\documentclass[12pt]{article}
\usepackage{amsmath}
\usepackage{latexsym}
\usepackage{amsfonts}
\usepackage[normalem]{ulem}
\usepackage{soul}
\usepackage{array}
\usepackage{amssymb}
\usepackage{extarrows}
\usepackage{graphicx}
\usepackage[backend=biber,
style=numeric,
sorting=none,
isbn=false,
doi=false,
url=false,
]{biblatex}

\usepackage{subfig}
\usepackage{wrapfig}
\usepackage{wasysym}
\usepackage{enumitem}
\usepackage{adjustbox}
\usepackage{ragged2e}
\usepackage[svgnames,table]{xcolor}
\usepackage{tikz}
\usepackage{longtable}
\usepackage{changepage}
\usepackage{setspace}
\usepackage{hhline}
\usepackage{multicol}
\usepackage{tabto}
\usepackage{float}
\usepackage{multirow}
\usepackage{makecell}
\usepackage{fancyhdr}
\usepackage[toc,page]{appendix}
\usepackage[hidelinks]{hyperref}
\usetikzlibrary{shapes.symbols,shapes.geometric,shadows,arrows.meta}
\tikzset{>={Latex[width=1.5mm,length=2mm]}}
\usepackage{flowchart}\usepackage[paperheight=11.0in,paperwidth=8.5in,left=1.0in,right=1.0in,top=1.0in,bottom=1.0in,headheight=1in]{geometry}
\usepackage[utf8]{inputenc}
\usepackage[T1]{fontenc}
\TabPositions{0.5in,1.0in,1.5in,2.0in,2.5in,3.0in,3.5in,4.0in,4.5in,5.0in,5.5in,6.0in,}

\renewcommand{\_}{\kern-1.5pt\textunderscore\kern-1.5pt}

\setlistdepth{9}
\renewlist{enumerate}{enumerate}{9}
		\setlist[enumerate,1]{label=\arabic*)}
		\setlist[enumerate,2]{label=\alph*)}
		\setlist[enumerate,3]{label=(\roman*)}
		\setlist[enumerate,4]{label=(\arabic*)}
		\setlist[enumerate,5]{label=(\Alph*)}
		\setlist[enumerate,6]{label=(\Roman*)}
		\setlist[enumerate,7]{label=\arabic*}
		\setlist[enumerate,8]{label=\alph*}
		\setlist[enumerate,9]{label=\roman*}

\renewlist{itemize}{itemize}{9}
		\setlist[itemize]{label=$\cdot$}
		\setlist[itemize,1]{label=\textbullet}
		\setlist[itemize,2]{label=$\circ$}
		\setlist[itemize,3]{label=$\ast$}
		\setlist[itemize,4]{label=$\dagger$}
		\setlist[itemize,5]{label=$\triangleright$}
		\setlist[itemize,6]{label=$\bigstar$}
		\setlist[itemize,7]{label=$\blacklozenge$}
		\setlist[itemize,8]{label=$\prime$}

\begin{document}
\begin{Center}
{\fontsize{14pt}{16.8pt}\selectfont \textbf{Artificial Intelligence in Quantitative Ultrasound Imaging: A Review}\par}
\end{Center}\par

\vspace{\baselineskip}
\begin{Center}
{\fontsize{11pt}{13.2pt}\selectfont Boran Zhou, Xiaofeng Yang and Tian Liu\par}
\end{Center}\par

\vspace{\baselineskip}
\begin{Center}
\textsuperscript{1}{\fontsize{11pt}{13.2pt}\selectfont Department of Radiation Oncology and Winship Cancer Institute, Emory University, Atlanta, GA\par}
\end{Center}\par

\vspace{\baselineskip}

\vspace{\baselineskip}

\vspace{\baselineskip}
{\fontsize{11pt}{13.2pt}\selectfont \textbf{Corresponding author:}\par}\par

{\fontsize{11pt}{13.2pt}\selectfont Tian Liu, PhD\par}\par

{\fontsize{11pt}{13.2pt}\selectfont Department of Radiation Oncology\par}\par

{\fontsize{11pt}{13.2pt}\selectfont Emory University School of Medicine\par}\par

{\fontsize{11pt}{13.2pt}\selectfont 1365 Clifton Road NE\par}\par

{\fontsize{11pt}{13.2pt}\selectfont Atlanta, GA 30322\par}\par

{\fontsize{11pt}{13.2pt}\selectfont Tel: (404)-778-1848\par}\par

{\fontsize{11pt}{13.2pt}\selectfont Fax: (404)-778-4139\par}\par

{\fontsize{11pt}{13.2pt}\selectfont E-mail: tian.liu@emory.edu\par}\par

 %%%%%%%%%%%%  Starting New Page here %%%%%%%%%%%%%%

\newpage

\vspace{\baselineskip}\begin{Center}
{\fontsize{11pt}{13.2pt}\selectfont \textbf{ABSTRACT}\par}
\end{Center}\par

\vspace{\baselineskip}
\begin{justify}
{\fontsize{11pt}{13.2pt}\selectfont Quantitative ultrasound (QUS) imaging is a reliable, fast and inexpensive technique to extract physically descriptive parameters for assessing pathologies. Despite its safety and efficacy, QUS suffers from several major drawbacks: poor imaging quality, inter- and intra-observer variability which hampers the reproducibility of measurements. Therefore, it is in great need to develop automatic method to improve the imaging quality and aid in measurements in QUS. In recent years, there has been an increasing interest in artificial intelligence (AI) applications in ultrasound imaging. However, no research has been found that surveyed the AI use in QUS. The purpose of this paper is to review recent research into the AI applications in QUS. This review first introduces the AI workflow, and then discusses the various AI applications in QUS. Finally, challenges and future potential AI applications in QUS are discussed. \par}
\end{justify}\par

\vspace{\baselineskip}

\vspace{\baselineskip}
\begin{justify}
{\fontsize{11pt}{13.2pt}\selectfont \textbf{\textit{Keywords:} }Quantitative ultrasound imaging (QUS); Artificial intelligence (AI); Deep learning (DL); Machine learning; Beamforming; Ultrasound elastography; Contrast enhanced ultrasound; Image analysis.

 %%%%%%%%%%%%  Starting New Page here %%%%%%%%%%%%%%

\newpage
\par}
\end{justify}\par

\section{Introduction}
{\fontsize{11pt}{13.2pt}\selectfont Compared with computed tomography (CT) and magnetic resonance imaging (MRI), ultrasound (US) imaging has key advantages including real-time imaging, non-ionizing nature and cost effectiveness. In addition, US is portable, requires no shielding, and utilizes conventional electrical power sources and is therefore well suited to point-of-care applications, especially in under-resourced settings. Quantitative ultrasound (QUS) imaging extracts physically interpretable, system-independent and uncorrelated parameters for acoustic interactions with tissue from radiofrequency (RF) signals, providing objective and quantitative measurement of tissue microstructure and pathological states [1]. This technique offers dedicated solutions based on signal- or image-derived quantitative measurements for evaluating a variety of diseases in cardiology, radiology and oncology. This technique is useful for screening and monitoring disease progression, image-guided intervention or therapy and improving the specificity of ultrasound imaging [2]. However, certain limitations exist in QUS, such as poor imaging quality caused by noise and artifacts, high dependence on abundant operator or diagnostician experience, and high inter- and intra-observer variability across different institutes and manufacturers’ US systems. In order to overcome these limitations, it is of great importance to develop advanced automatic US signal or image analysis methods to make US diagnosis and/or assessment, as well as image-guided interventions/therapy, more objective, accurate and intelligent. \par}\par

\begin{justify}
{\fontsize{11pt}{13.2pt}\selectfont Enhanced with increasingly powerful hardware, open source software and available public datasets, artificial intelligence (AI) has been extensively applied in many fields, such as computer vision, natural language processing, \textit{etc}. AI has been successfully conducted in medical image analysis and image-based assessment of disease states, illustrating that AI is a cutting-edge tool for medical image analysis, acting as a tool for the rapid, accurate assessment of pathological structures and functions [3]. Machine learning (ML) is a subfield of AI where computers are trained to perform tasks without explicit programming [4, 5]. Representation learning is a subfield of ML where no feature engineering is used; instead, the computer learns the features by which to classify or repress the provided data. Deep learning (DL) is a subfield of representation learning where the learned features are compositional or hierarchical. \par}
\end{justify}\par

\begin{justify}
{\fontsize{11pt}{13.2pt}\selectfont Numerous review articles focused on the AI applications in the medical ultrasound, either in computer-aided diagnostic systems or ultrasound imaging [5-8]. However, no research has been found that surveyed the AI applications in QUS. This paper systematically reviews and aims to provide a comprehensive understanding of AI applications in QUS. The rest of the paper is organized as follows: In Section 2, we introduce the AI workflow. In Section 3, we discuss in detail the AI applications in QUS. In Section 4, we present challenges and potential AI applications in QUS. \par}
\end{justify}\par

\begin{justify}
{\fontsize{11pt}{13.2pt}\selectfont   \par}
\end{justify}\par

\section{Artificial Intelligence}
\begin{justify}
{\fontsize{11pt}{13.2pt}\selectfont Artificial intelligence (AI) is a subfield of computer science devoted to creating systems to perform tasks ordinarily requiring human intelligence. Machine learning (ML) is a subfield of AI, aiming to enable computer to conduct certain tasks based on previous experience. ML methods can be categorized into supervised, unsupervised or semi-supervised learning based on whether the collected dataset for training is fully labeled, unlabeled or partially labeled. Deep learning (DL), a subset of ML and a data-driven methodology, extracts multiple levels of nonlinear features from datasets and aims to make inference based on the learned data. ML\ algorithms include Naïve Bayes, support vector machine (SVM), random forest, Artificial Neural Network (ANN), etc.  Naïve Bayes algorithm is a probabilistic classifier via calculating the maximum posterior probability based on the prior probability and the observed likelihood in the training set [9]. SVM projects the data from a low dimensional to a high dimensional feature space so as to improve linear separability via developing a super hyperplane with the largest margin between positive and negative samples in the feature space [10]. As an ensemble learning algorithm, random forest aims to train a strong learner by combining the predictions of multiple weak learners. This algorithm is effective in reducing the overfitting and robust to situations in which the discrete values are in the feature space [11].\ \ \  \  \par}
\end{justify}\par

\begin{justify}
{\fontsize{11pt}{13.2pt}\selectfont Artificial neural network (ANN) consists of artificial neurons and their connections which have weights to reflect inter-dependencies between neurons. Each neuron receives multiple inputs and its activation is determined by the overall input. The activated neuron outputs a signal. The weights associated with the neurons are adaptively adjusted by a learning algorithm, such as back propagation, to make the predictions as close to the targets as possible. Despite its excellent performance in various fields, certain limitations are associated with ANN, such as decrease in the local minimum during optimization and overfitting. Deep neural network (DNN), which consists of multiple layers of neurons, is used to overcome these limitations and capable of handling large and complex datasets. The first layer of DNN is the input layer while the final layer is the output layer for tasks such as classification, segmentation or regression. The layers between input and output layers are hidden layers which extract multiple levels of abstract features from input data. The number of neurons in the hidden layers determines the learning complexity. The associated weights used by DNN are usually randomly initialized and optimized by algorithms such as gradient descent to find a local minimum. The training dataset is fed into the DNN and a loss function\ between\ the prediction and the label is minimized for updating the weights. Before training the model, the data is usually preprocessed and augmented.   \par}
\end{justify}\par

\subsection{Data augmentation and feature selection}
\begin{justify}
{\fontsize{11pt}{13.2pt}\selectfont In the field of medical imaging, access to data is heavily protected due to privacy regulations, making data insufficient and hindering the inference accuracy of trained model. The models trained with small datasets usually do not generalize well and severely suffer from overfitting. Via increasing the amount of data based only on the training dataset, image augmentation is essential to teach the model the desired invariance and used to effectively reduce overfitting [12]. Image augmentation techniques include geometric and color augmentations, such as rotation, scaling, flipping, translation, cropping and change in the color palette of the image. Various packages have been developed and are publicly available for image augmentation: such as imgaug, Albumentations, Augmentor, Image preprocessing in Keras, torchvision.transform in Pytorch, etc [13-17]. \par}
\end{justify}\par

\begin{justify}
{\fontsize{11pt}{13.2pt}\selectfont However, these augmentation techniques do not take variations from different imaging protocols or variations in size, shape or location of specific pathology into account. Generative Adversarial Nets (GAN) is a neural network model where two networks are trained simultaneously with one model (generator) for generating artificial data and the other (discriminator) focused on discriminating artificial data from actual data [18]. The discriminator encourages the generator to generate realistic data through penalizing false data via learning. This technique provides a more generic solution and has been extensively utilized in exploring and discovering the underlying structure of training data and augmenting training images, overcoming the privacy issues related to medical image data and dealing with the insufficient number of positive pathological cases [19]. \par}
\end{justify}\par

\begin{justify}
{\fontsize{11pt}{13.2pt}\selectfont Irrelevant or partially relevant features can negatively impact model performance [20]. Feature selection, a process of selecting a subset of relevant features to use in model training, is one of the core algorithms in machine learning, significantly reducing overfitting and training time as well as improving accuracy. Feature selection algorithms can be classified as three methods: filter, wrapper and embedded methods [21]. Filter methods assign a score to each feature via a statistical method based on its correlation with the output. These features are then selected or removed from the dataset based on the score. Wrapper methods use a predictive model to evaluate a combination of features and assign a score based on the model accuracy. Embedded method is a combination of filter and wrapper methods. It selects features best contribute to the model accuracy while the model is being constructed. \par}
\end{justify}\par

\subsection{Supervised learning}
\begin{justify}
{\fontsize{11pt}{13.2pt}\selectfont In supervised learning, the computer learns to classify data by providing a training dataset of labeled data and minimize the error with known labels. Its ability for prediction is then tested on an unseen dataset. Convolutional neural network (CNN) and recurrent neural network (RNN) are the two of the most widely used architectures. \par}
\end{justify}\par

\begin{justify}
{\fontsize{11pt}{13.2pt}\selectfont A typical CNN usually consists of multiple convolutional layers, pooling layers, batch normalization layers, dropout layers, fully connected layers. A convolutional layer is used to extract multiple levels of visual features via multiplication of trainable kernels with local neighbor of a pixel across the input feature maps. The convolutional layer is composed of multiple kernels and generates the same number of feature maps. Multiple convolutional layers can be used to extract hierarchical abstract features [22]. Pooling layers are often used to subsample the feature maps so as to capture a large field of view and promote spatial invariance of the network. Max or average pooling is usually used. Convolutional and pooling layers are usually stacked in various orders to build a deep CNN architecture. Batch normalization layer is used to reduce covariance shift in the training dataset. Weight regularization and dropout layers are mostly used to reduce overfitting. The fully connected layers are incorporated at the end of the network to output the final results. Various kinds of network architectures have been developed to improve the performance of deep CNN. U-Net, which uses symmetrical contractive and expansive paths with skip connections between them, has widely been used for structure segmentation [23]. In order to alleviate gradient vanishing or explosion in training deep neural networks, a residual network (ResNet) was proposed via learning residual functions [24]. Moreover, a densely connected convolutional network (DenseNet) was proposed via connecting each layer to every other layer [25]. \par}
\end{justify}\par

\begin{justify}
{\fontsize{11pt}{13.2pt}\selectfont Recurrent neural network (RNN), in which the interconnections between artificial neurons form a unidirectional cycle, is a type of ANN specialized for learning dynamic temporal data. In RNN, a hidden state,  \( h_{t} \) , at time  \( t \)  that is the output of a nonlinear mapping from its input, and the previous state  \( h_{t-1} \) , \par}
\end{justify}\par

\begin{FlushRight}
 \( h_{t}= \sigma  \left( Wx_{t}+Rh_{t-1}+b \right) ~ \) {\fontsize{11pt}{13.2pt}\selectfont \ \ \ \ \ \ \ \ \ \ \ \ \ \ \ \ \ \ \ \ \ \ \ \ \ \ \ \ \ \ \ \ \ \ \ \ \ \ \ \ \ \ \  (1)\par}
\end{FlushRight}\par

\begin{justify}
{\fontsize{11pt}{13.2pt}\selectfont where\ the weights   \( W~ \) and  \( R \)  are shared over time,  \( b \)  is a bias [26]. \par}
\end{justify}\par

\begin{justify}
{\fontsize{11pt}{13.2pt}\selectfont \tab RNN has widely been used in the fields of natural language processing or text recognition [26, 27]. However, this technique is seldomly used in the domain of medical imaging. \par}
\end{justify}\par

%\begin{enumerate}[label*={\fontsize{11pt}{11pt}\selectfont \textbf{\arabic*.}}]
	%\item {\fontsize{11pt}{13.2pt}\selectfont \textbf{Reinforcement learning }\par}\par
\subsection{Reinforcement learning}
\begin{justify}
{\fontsize{11pt}{13.2pt}\selectfont Reinforcement learning (RL) is a kind of machine learning which performs the prediction of the best actions to take considering its current environmental state. \  This technique basically involves sequentially making decisions and can be modeled\ as a Markov decision process in which an artificial agent is trained to maximize the cumulative expected rewards given a set of states and actions.  Q-learning, a model-free RL algorithm, aims to model the action-reward relationship with a Q function. In the field of imaging processing, the CNN is usually used to model the Q function which can be trained via supervised learning [28]. \par}
\end{justify}\par

	%\item {\fontsize{11pt}{13.2pt}\selectfont \textbf{Unsupervised} \textbf{learning}\par}\par
\subsection{Unsupervised learning}
\begin{justify}
{\fontsize{11pt}{13.2pt}\selectfont In contrast to supervised learning which requires human annotations and is correspondingly labor-intensive and expensive, unsupervised learning uses unlabeled data to group the input data into multiple clusters based on similarities among the data. It includes Auto-Encoders (AE), restricted Boltzmann’s machines (RBMs)/deep belief networks (DBNs), etc [29, 30]. AE, which usually consists of an encoder which encodes the input into a low-dimensional latent state space and a decoder which decodes the output from the low-dimensional latent space, is a nonlinear feature extractor and used for representation learning [31]. Variants of AE, such as sparse auto-encoders (SAEs) and denoising auto-encoders (DAEs) have been developed [32, 33]. Regularization and sparsity constraints are applied in SAEs in order to strengthen the training and prevent an autoencoder from learning an identity function. With the introduction of the denoising criterion, DAEs are locally trained to denoise corrupted versions of inputs. RBM consists of a visible (input) layer and a hidden layer without intra-layer communication. Two-layer neural nets constitute the building blocks of the DBNs. A DBN consists of a visible layer and multiple hidden layers with the top two layers as an RBM and the lower layers as a sigmoid belief network. Including a linear classifier on top of the DBN, a specific task such as object detection, segmentation and classification can be performed. \par}
\end{justify}\par

%\end{enumerate}\subsubsection{Transfer learning}
\subsection{Transfer learning}
\begin{justify}
{\fontsize{11pt}{13.2pt}\selectfont A deep CNN model usually consists of tens of millions of parameters to train, requiring a large number of annotated images. Collection and annotation of large number of medical images is a big hurdle for deep learning (DL) applications in the field of medical imaging. Transfer learning, in which models can be fine-tuned from pre-trained models, has widely been adopted [34-36]. For transfer learning, its hypothesis is that despite the dissimilarity between natural and medical images, the deep CNN models trained on the large scale annotated natural image dataset may be transferrable to enable the recognition in the medical image [34]. In the field of medical imaging, strategies using a pre-trained network to extract features or fine-tuning a pre-trained network on medical images have widely been utilized [37, 38].\par}
\end{justify}\par

\section{AI applications in QUS}

\vspace{\baselineskip}
\begin{justify}
{\fontsize{11pt}{13.2pt}\selectfont The papers included in this review were derived from an initial search of the key words $``$artificial intelligence machine deep learning quantitative ultrasound imaging$"$  in PubMed, Web of Science and Google Scholar in the last thirteen years. Once a considerable amount of papers was collected, the relevant sources from these papers were also added to a manual review for a more comprehensive review of the field of AI in QUS, as summarized in Figure 1. We totally collected over 150 papers that are closely related to machine or deep learning in QUS. Most of these works were published between 2016 and 2019. The number of papers has grown dramatically over the past few years. \par}
\end{justify}\par

\begin{justify}
{\fontsize{11pt}{13.2pt}\selectfont Principal applications of AI in QUS include beamforming, elastography, 3d freehand US, US image analysis. These applications are surveyed in the following subsections. \par}
\end{justify}\par

%%%%%%%%%%%%%%%%%%%% Figure/Image No: 1 starts here %%%%%%%%%%%%%%%%%%%%

\begin{figure}[H]
	\begin{Center}
		\includegraphics[width=6.5in,height=2.49in]{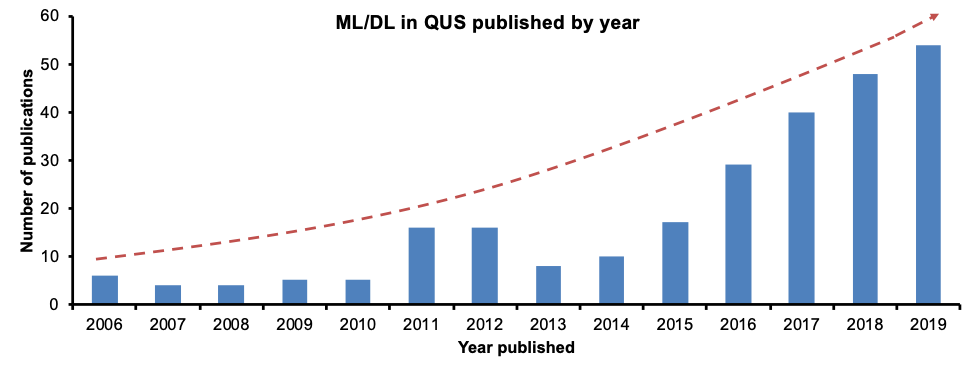}
	\end{Center}
\end{figure}

%%%%%%%%%%%%%%%%%%%% Figure/ No: 1 Ends here %%%%%%%%%%%%%%%%%%%%

\par

\begin{Center}
{\fontsize{11pt}{13.2pt}\selectfont \textbf{Figure 1.} Number of AI applications in QUS related papers published since 2006.\par}
\end{Center}\par

\vspace{\baselineskip}
\subsection{Beamforming}
\begin{justify}
{\fontsize{11pt}{13.2pt}\selectfont Multi-line acquisition (MLA) has been used in cardiac ultrasound imaging to obtain a high frame rate so as to accurately capture rapid motion. Vedula \textit{et al}. proposed to train two encoder-decoder neural networks for the I and Q signals and apply this technique to the whole transmit-receive pipeline [39]. The unfocused multi-line channel data as input was fed into the network and the output was beamformed image which was as close to the image from single-line acquisition as possible via minimizing the  \( l_{1} \)  distance. They demonstrated that jointly training the transmit patterns with receive beamforming greatly improve the image quality. In order to improve the imaging quality in MLA, this group then used this framework to removal artifacts in time-delayed and phase-rotated element-wise I/Q data in MLA [40]. This network contains both the interpolation and the apodization stages. They demonstrated that this technique may be able to\ substitute the conventional ultrasound MLA correction.  Other DNNs, such as stacked autoencoders, encoder-decoder architectures and fully convolutional networks (FCN) were used for mapping pre-delayed channel data to beamformed images [41-45]. Yoon \textit{et al}. proposed to use CNN to interpolate the missing RF data by using redundancy in the receiver-transmit plane [46]. The CNN was used in the receive-transmit or receive-scan line domains to explore the redundancy in the RF domain. One advantage of this technique is that it does not need any hardware change and can be easily implemented onto any B-mode ultrasound system or transducer. Luchies \textit{et al}. used DNN for suppressing off-axis scattering in ultrasound channel data [47]. Discrete Fourier-transformation (DFT) was first used on the channel data to obtain in-phase and quadrature components at a frequency and a depth. The array responses were fed as input into the DNN and the frequency spectra as output were used to generate a beamformed RF signal at that depth via inverse discrete Fourier-transformation (IDFT) and summation across the array.  \par}
\end{justify}\par

\subsection{Elastography}
\begin{justify}
{\fontsize{11pt}{13.2pt}\selectfont Ultrasound elastography has widely been used for non-invasive assessment of tissue mechanical properties which are usually regarded as biomarkers for pathological processes, providing qualitative and quantitative information which can be used for diagnostic purposes. Ultrasound elastography has been successfully applied to grade chronic liver disease [48], assess breast lesions [49], thyroid nodules [50], prostate cancer [51], interstitial lung disease [52-55], etc. However, ultrasound elastography suffers from inter- and intra-observer variability and hampers the reproducibility of measurements. Moreover, assumptions of linearity, incompressibility, isotropic, elasticity for simplifying analysis and interpretation of measurements violate models of soft tissue as complex and heterogeneous material which have both elastic and viscous properties. These limitations illustrate the potential of using AI in ultrasound elastography. \par}
\end{justify}\par

\begin{justify}
{\fontsize{11pt}{13.2pt}\selectfont Wu \textit{et al}. used the separable convolution to extract and concatenate the hybrid features of pre- and post-compression RF data in the displacement estimation stage. [56]. In the strain prediction stage, they used another convolution network with three layers to extract the high-level semantic information of the tissue displacement for strain map prediction. This technique resulted in better signal-to-noise ratio (SNR) and contrast-to-noise ratio (CNR). In the quasi-static ultrasound elastography, it usually requires explicit model for strain reconstruction. However, this technique suffers weak model adaptation, low efficiency and high user experience requirement. In order to address this issue, this group later proposed the learning-using-privileged-information paradigm with causality in the network [57]. This framework was based on the DNN for implicit strain reconstruction to infer the map from the RF data to tissue strain field. Ahmed \textit{et al}. used DNN to reconstruct ultrasound shear wave elastography (USWE) from tackled tissue displacement data at different time instances induced by a single acoustic radiation force [58]. A localizer was used to locate the lesion and RNN was used to extract temporal correlations from wave patterns at various time frames. Finally, an estimator was used to reconstruct shear modulus of lesion from concatenated outputs of localizer and RNN. However, shear wave imaging requires high-end ultrasound hardware and high power. Given longitudinal wave speed provides similar diagnostic capacities compared to shear wave imaging, Feigin \textit{et al}. used a CNN for measuring single sided pressure-wave sound speed from channel data [59]. This technique was validated on simulated data and shown that it was possible to invert for longitudinal wave speed measurement for soft tissue at high frame rates. \par}
\end{justify}\par

\subsection{3D freehand US imaging}
\begin{justify}
{\fontsize{11pt}{13.2pt}\selectfont The 3D freehand US imaging has advantages such as low cost, high volume quality and field of view. However, given the errors from localization sensors and low calibration accuracy as well as delays at the reconstruction, it is difficult and cumbersome to manually capture a 3D US volume with this technique [60]. Prevost \textit{et al}. proposed an end-to-end technique based on CNN to learn the relative 3D translations and rotations from a pair of 2D ultrasound images and correspondingly to predict the ultrasound transducer motion [61]. Provided with in-plane displacement, the network input consists of 4 channels: the two successive images and the two components of the estimated vector field. Based on the measurements from an inertial measurement unit (IMU), this group also reconstructed 3D ultrasound volumes from sequences of free-hand images obtained with 2D transducer [62, 63]. The input of the architecture are pair of frames, optical flow vector field and the IMU measurements while the output is a 6-dimensional vector representing the translations and rotations. A trajectory reconstruction algorithm was then implemented to chain all the estimated transformations. \par}
\end{justify}\par

\subsection{US image analysis}
\subsubsection{Classification}
%\begin{enumerate}[label*={\fontsize{11pt}{11pt}\selectfont \arabic*.}]
	%\item \begin{enumerate}[label*={\fontsize{11pt}{11pt}\selectfont \arabic*.}]
	%\item \begin{enumerate}[label*={\fontsize{11pt}{11pt}\selectfont \arabic*.}]
	%\item \begin{enumerate}[label*={\fontsize{11pt}{11pt}\selectfont \arabic*.}]
	%\item {\fontsize{11pt}{13.2pt}\selectfont Temporal enhanced ultrasound \par}\par

%\vspace{\baselineskip}
\paragraph{Temporal enhanced ultrasound}
\begin{justify}
{\fontsize{11pt}{13.2pt}\selectfont Temporal-enhanced ultrasound extracts information from the temporal sequence of backscattered US RF data of the region of interest (ROI) [64, 65]. In 2016, Azizi \textit{et al}. used a deep belief network (DBN) to learn the high-level latent features and a SVM classifier to differentiate cancerous versus benign tissue [66]. Given RF data is not available on all commercial US systems, this group then used transfer learning from RF to B-mode features for detecting prostate cancer [67]. Domain adaptation and transfer learning techniques were combined to train a tissue classification model on RF data (source domain) and B-mode data (target domain). The same group later incorporated the DL back-end into a platform for real time prostate biopsy [68-70]. RNN with long short-term memory (LSTM) cells was used for prostate cancer classification while residual neural networks and dilated CNN were used for prostate segmentation. \par}
\end{justify}\par

\paragraph{Echocardiography}
	%\item {\fontsize{11pt}{13.2pt}\selectfont Echocardiography \par}\par

\begin{justify}
{\fontsize{11pt}{13.2pt}\selectfont Given views in echocardiography exhibit subtle difference from each other, the first nontrivial step to evaluate an echocardiogram is to identify the view. Table 1 shows a list of selected references that used AI in echocardiography classification. Park \textit{et al}. used a multi-class logist-boosting algorithm for cardiac view classification [71]. Four standard cardiac views were classified: apical four chamber and apical two chamber, parasternal long axis and parasternal short axis. They firstly built a left ventricle (LV) detector per view to obtain local information for a binary classification and then employed the view-specific knowledge to incorporate the local information for global view classification. The classification accuracies over 96$\%$  were obtained in both training and testing datasets. Chykeyuk \textit{et al}. used a relevance vector machine (RVM) classifier, which is a probabilistic extension of SVM and provides posterior probabilities of test data, for cardiac wall motion classification of stress echocardiography based on features from rest and stress sequences [72]. 19 out of 30 features were selected via a feature selection technique. The overall accuracy of global wall motion classification was 93.02$\%$ , outperforming the Hidden Markev Model. Sengupta \textit{et al}. developed an associative memory classifier (AMC), which associates one set of vectors (input patterns) with another set of vectors (output patterns) using the associative memory units, to differentiate constrictive pericarditis from restrictive cardiomyopathy based on speckle tracking echocardiography (STE) [73]. The STE data was spatially and temporally normalized and binned. This model achieved area under the curve (AUC) of 89.2$\%$  which was then improved to 96.2$\%$  with addition of 4 echocardiographic variables. Samad \textit{et al}. used a random forest and a logistic regression models to predict the survival rate based on echocardiography-derived measurements combined with clinical electronic health record data [74]. It showed that the random forest model reached 96$\%$  of the maximum prediction accuracy with 6 variables derived from echocardiography and outperformed the logistic regression model. \par}
\end{justify}\par

\begin{justify}
{\fontsize{11pt}{13.2pt}\selectfont Madani \textit{et al}. adopted a CNN model to simultaneously classify 15 standard views: parasternal long axis, right ventricular inflow, basal short axis, short axis at mid or mitral level, apical four-chamber, apical five chamber, apical two chamber, apical three chamber [75]. Differences in zoom, depth, focus, sector width, gain of images for each view were included. The training dataset was augmented by applying rotation, scaling, subtraction by the mean value, width and height shift, zoom, shearing, and vertical/horizontal flips. VGG-16 network was used for this task to achieve an overall accuracy of 97.8$\%$  among 12 video views in the testing dataset. The same group in 2018 developed a semi-supervised GAN model which can learn from both labeled and unlabeled data in a generalizable fashion [76]. View segmentation was performed prior to view classification. The effect of image resolution on model performance and computational cost was assessed by varying image resolution. The obtained results showed that the model was able to achieve an accuracy greater than 80$\%$  with less than 4$\%$  of the data. Gao \textit{et al}. developed a fused DL framework incorporating both spatial and temporal information sustained by the video clips for eight viewpoints classification [77]. Dezaki \textit{et al}. developed a deep residual neural network (ResNet) combined with RNN for automatic recognition of cardiac cycle phase [78]. The ResNet extracts hierarchical features from the individual echocardiogram frame while RNN model the temporal dependencies between sequential frames. \par}
\end{justify}\par

{\fontsize{11pt}{13.2pt}\selectfont \textbf{Table 1.} Overview of AI applications in echocardiography classification. \par}\par
%%%%%%%%%%%%%%%%%%%% Table No: 1 starts here %%%%%%%%%%%%%%%%%%%%
\begin{table}[H]
 			\centering
\begin{tabular}{p{0.3in}p{0.36in}p{2.18in}p{2.86in}}
\hline
%row no:1
\multicolumn{1}{|p{0.3in}}{{\fontsize{10pt}{12.0pt}\selectfont Ref}} & 
\multicolumn{1}{|p{0.36in}}{{\fontsize{10pt}{12.0pt}\selectfont Year}} & 
\multicolumn{1}{|p{2.18in}}{{\fontsize{10pt}{12.0pt}\selectfont Network}} & 
\multicolumn{1}{|p{2.86in}|}{{\fontsize{10pt}{12.0pt}\selectfont $\#$  of samples in training/testing datasets}} \\
\hhline{----}
%row no:2
\multicolumn{1}{|p{0.3in}}{{\fontsize{10pt}{12.0pt}\selectfont [71]}} & 
\multicolumn{1}{|p{0.36in}}{{\fontsize{10pt}{12.0pt}\selectfont 2007}} & 
\multicolumn{1}{|p{2.18in}}{{\fontsize{10pt}{12.0pt}\selectfont Logist-boosting}} & 
\multicolumn{1}{|p{2.86in}|}{{\fontsize{10pt}{12.0pt}\selectfont 1080/223 sequences}} \\
\hhline{----}
%row no:3
\multicolumn{1}{|p{0.3in}}{{\fontsize{10pt}{12.0pt}\selectfont [72]}} & 
\multicolumn{1}{|p{0.36in}}{{\fontsize{10pt}{12.0pt}\selectfont 2011}} & 
\multicolumn{1}{|p{2.18in}}{{\fontsize{10pt}{12.0pt}\selectfont RVM}} & 
\multicolumn{1}{|p{2.86in}|}{{\fontsize{10pt}{12.0pt}\selectfont 173 subjects, 5-fold CV}} \\
\hhline{----}
%row no:4
\multicolumn{1}{|p{0.3in}}{{\fontsize{10pt}{12.0pt}\selectfont [73]}} & 
\multicolumn{1}{|p{0.36in}}{{\fontsize{10pt}{12.0pt}\selectfont 2016}} & 
\multicolumn{1}{|p{2.18in}}{{\fontsize{10pt}{12.0pt}\selectfont AMC}} & 
\multicolumn{1}{|p{2.86in}|}{{\fontsize{10pt}{12.0pt}\selectfont 94 subjects, 10-fold CV}} \\
\hhline{----}
%row no:5
\multicolumn{1}{|p{0.3in}}{{\fontsize{10pt}{12.0pt}\selectfont [74]}} & 
\multicolumn{1}{|p{0.36in}}{{\fontsize{10pt}{12.0pt}\selectfont 2018}} & 
\multicolumn{1}{|p{2.18in}}{{\fontsize{10pt}{12.0pt}\selectfont Random forest, logistic regression}} & 
\multicolumn{1}{|p{2.86in}|}{{\fontsize{10pt}{12.0pt}\selectfont 171510 subjects, 10-fold CV}} \\
\hhline{----}
%row no:6
\multicolumn{1}{|p{0.3in}}{{\fontsize{10pt}{12.0pt}\selectfont [75]}} & 
\multicolumn{1}{|p{0.36in}}{{\fontsize{10pt}{12.0pt}\selectfont 2018}} & 
\multicolumn{1}{|p{2.18in}}{{\fontsize{10pt}{12.0pt}\selectfont CNN}} & 
\multicolumn{1}{|p{2.86in}|}{{\fontsize{10pt}{12.0pt}\selectfont 200000/20000 images}} \\
\hhline{----}
%row no:7
\multicolumn{1}{|p{0.3in}}{{\fontsize{10pt}{12.0pt}\selectfont [76]}} & 
\multicolumn{1}{|p{0.36in}}{{\fontsize{10pt}{12.0pt}\selectfont 2018}} & 
\multicolumn{1}{|p{2.18in}}{{\fontsize{10pt}{12.0pt}\selectfont GAN}} & 
\multicolumn{1}{|p{2.86in}|}{{\fontsize{10pt}{12.0pt}\selectfont 645/77 subjects}} \\
\hhline{----}
%row no:8
\multicolumn{1}{|p{0.3in}}{{\fontsize{10pt}{12.0pt}\selectfont [77]}} & 
\multicolumn{1}{|p{0.36in}}{{\fontsize{10pt}{12.0pt}\selectfont 2017}} & 
\multicolumn{1}{|p{2.18in}}{{\fontsize{10pt}{12.0pt}\selectfont CNN, RNN}} & 
\multicolumn{1}{|p{2.86in}|}{{\fontsize{10pt}{12.0pt}\selectfont 280/152 images}} \\
\hhline{----}
%row no:9
\multicolumn{1}{|p{0.3in}}{{\fontsize{10pt}{12.0pt}\selectfont [78]}} & 
\multicolumn{1}{|p{0.36in}}{{\fontsize{10pt}{12.0pt}\selectfont 2017}} & 
\multicolumn{1}{|p{2.18in}}{{\fontsize{10pt}{12.0pt}\selectfont ResNet, RNN}} & 
\multicolumn{1}{|p{2.86in}|}{{\fontsize{10pt}{12.0pt}\selectfont N/A}} \\
\hhline{----}

\end{tabular}
 \end{table}

%%%%%%%%%%%%%%%%%%%% Table No: 1 ends here %%%%%%%%%%%%%%%%%%%%

\begin{justify}
{\fontsize{11pt}{13.2pt}\selectfont $\ast$ N/A: not available. \par}
\end{justify}\par

\paragraph{Contrast-enhanced ultrasound}
	%\item {\fontsize{11pt}{13.2pt}\selectfont Contrast-enhanced ultrasound \par}\par

\begin{justify}
{\fontsize{11pt}{13.2pt}\selectfont Contrast enhanced ultrasound (CEUS) utilizes an intravenous injection of gas-filled microbubble as contrast agent to support ultrasound imaging of micro-vascularization. The spatial and temporal dynamic patterns of the microbubble were recorded in the CEUS image sequences. This technique can be used to quantify perfusion in terms of the contrast uptake (or time-intensity-curve (TIC)) in a ROI outlining a tumor, detecting changes in tumor vascularity and monitoring tumor response to therapy in radiation oncology. The limitations of CEUS include operator dependence, motion effect and requirements of suitable acoustic window. AI may provide advanced image processing technique to alleviate these limitations. \par}
\end{justify}\par

\begin{justify}
{\fontsize{11pt}{13.2pt}\selectfont Wu \textit{et al}. used 12 grayscale and Doppler features for training a logistic regression model in classifying triple-negative breast cancer [79]. The quantitative features from grayscale image were margin sharpness, margin echogenicity difference, angular variance in margin, depth-to-width ratio, axis ratio, tortuosity, circularity, radius variation, and elliptically normalized skeleton. The quantitative features from color Doppler ultrasound were vascular indices of fractional area of flow in the lesion, mean flow velocity in the lesion, and flow volume in the lesion. The AUCs were 0.85 and\ 0.65 for grayscale and color Doppler features and it increased to 0.88 with sensitivity of 86.96$\%$  and specificity of 82.91$\%$  when the grayscale and color Doppler features were combined.  Mean values of the tumor vascular features in terms of vessel-to-volume ratio, number of vascular trees, total vessel length, the longest path length, number of bifurcations, and vessel diameter were computed based on 3D power Doppler ultrasound images with a 3D thinning algorithm to narrow down vessels to skeletons [49]. These values were fed as input to a multilayer perceptron (MLP) neural network to classify malignant or benign breast mass tumors. The obtained AUC values of the six features were 0.84, 0.87, 0.87, 0.82, 0.84 and 0.75, respectively. Based on combining all six features, the AUC value was increased to 0.92. \par}
\end{justify}\par

\begin{justify}
{\fontsize{11pt}{13.2pt}\selectfont Wu \textit{et al}. used DBNs for classifying liver masses based on CEUS images [80]. The dynamic CEUS videos of hepatic perfusion were retrieved and TICs were extracted from the videos using sparse non-negative matrix factorization. These TICs were used as input to classify benign and malignant focal liver lesions using a three-layer DBN. The highest accuracy of 86.36$\%$  was achieved using this technique which outperform other methods such as k-nearest neighbors (KNN), SVM, back propagation net (BPN). Guo \textit{et al}. proposed a two-stage multi-view learning framework based on CEUS for diagnosing liver tumors [81-83]. This technique only adopts three typical CEUS images selected from the arterial phase, portal venous phase and late phase. The deep canonical correlation analysis (DCCA) was performed on three image pairs and generated total six-view features. In the second stage, these multi-view features were fed into a multiple kernel learning (MKL) based classifier for diagnosis. The images obtained from CEUS were segmented and registered. The model-based features related to contrast perfusion and dispersion were extracted from the CEUS videos. This DCCA-MKL algorithm achieved the AUC value of 0.953 with sensitivity of 93.56 $ \pm $  5.9$\%$  and specificity of 86.89 $ \pm $  9.38$\%$ . In 2018, Feng \textit{et al}. proposed a 3D CNN to extract spatial-temporal features from CEUS sequential images for prostate cancer detection [84]. The frames of the CEUS videos were split into a set of 3D tensor samples. The targeted CEUS with anti-PSMA (prostate specific membrane antigen) agent increased the sensitivity and specificity for prostate cancer detection using DL compared to the non-targeted samples. The results showed this technique achieved over 91$\%$  in specificity and 90$\%$  in average accuracy. \par}
\end{justify}\par

\paragraph{Ultrasound elastography}
	%\item {\fontsize{11pt}{13.2pt}\selectfont Ultrasound elastography \par}\par

\begin{justify}
{\fontsize{11pt}{13.2pt}\selectfont 
 shows a list of selected references that used AI in ultrasound elastography classification. To improve the diagnosis accuracy for hepatitis B staging, Chen \textit{et al}. used eleven real-time tissue elastography images from subjects who underwent liver biopsies to train four classifiers, SVM, Naïve Bayes, Random Forest and K-Nearest [85]. Random forest classifier obtained the highest accuracy of 0.8287 with sensitivity of 0.8941 and specificity of 0.6499 among the four ML algorithms. In 2017, Gatos \textit{et al}. proposed to use a SVM classifier to grade the chronic liver disease based on the features from ultrasound shear wave elastography (SWE) [86]. An inverse mapping procedure was performed to obtain the stiffness value from the RGB color map. Then a clustering procedure was conducted to classify the chronic liver disease based on the stiffness values. The obtained accuracy was 87.3$\%$  with a sensitivity of 93.5$\%$  and specificity of 81.2$\%$ . Sehgal \textit{et al}. used Naïve Bayes and logistic regression models for classifying breast masses based on the grayscale and shape features from breast sonography as well as patient age and mammographic category of lesion based on Breast Imaging Reporting and Data System (BI-RADS) [87]. The obtained AUCs were 0.902 $ \pm $  0.023 and 0.865 $ \pm $  0.027 for logistic regression and Naïve Bayes, respectively. Moreover, it showed that the combined use of logistic regression and Naïve Bayes models demonstrated a reduction in biopsies by 48$\%$  with false negative rate of 6.4$\%$ . Gatos \textit{et al}. used a pre-trained GoogLeNet architecture for staging liver fibrosis based on the SWE images with temporal stability masks, showing accuracy (ranging from 82.5$\%$  to 95.5$\%$ ) for various chronic liver disease stage combinations [88]. Zhang \textit{et al}. adopted the point-wise gated Boltzmann machine and RBM for classifying breast tumor based on SWE images with an accuracy of 93.4$\%$ , a sensitivity of 88.6$\%$ , a specificity of 97.1$\%$  and AUC of 0.947 [89]. The same group later used a deep polynomial network (DPN) for breast tumor classification based on segmented dual-modal image features from both SWE and B-mode ultrasound, showing a sensitivity of 97.8$\%$ , a specificity of 94.1$\%$ , an accuracy of 95.6$\%$  and AUC of 0.961 [89-91]. \par}
\end{justify}\par

\begin{justify}
{\fontsize{11pt}{13.2pt}\selectfont Limitations of shear wave imaging (SWI) on prostate include small ROI size, slow frame rate, delays to achieve image stabilization, signal attenuation. Shi \textit{et al}. proposed a stacked DPN (S-DPN) algorithm to classify the prostate tumor based on a small dataset (70 prostate ultrasound elastography images with 42 benign masses and 28 malignant tumors) [92]. This model achieved a classification accuracy of 90.28 $ \pm $  2.78$\%$  with sensitivity of 84.14 $ \pm $  3.24$\%$  and specificity of 93.49 $ \pm $  4.45$\%$ . Gao \textit{et al}. investigated the plantar fasciitis classification based on SWE images using a deep Siamese framework with multitask learning and transfer learning (DS-MLTL) frameworks [93]. Discriminative visual features and effective recognition functions were learned. This method achieved a favorable accuracy of 85.09 $ \pm $  6.67$\%$ . Zhou \textit{et al}. used the wave speeds obtained from lung ultrasound surface wave elastography to predict mass density of superficial lung tissue using a densely neural network [94]. Synthetic data was generated for training and this model was validated on a sponge experiment. The obtained results showed that the accuracy was 0.992 in the test dataset. This group later used this technique on the in vivo datasets with 77 healthy subjects and patients with interstitial lung disease with the measurements from pulmonary function test as the ground truth [95].\  They obtained an accuracy of 0.89 in the testing dataset. \par}
\end{justify}\par

\vspace{\baselineskip}
\vspace{\baselineskip}
\vspace{\baselineskip}
\vspace{\baselineskip}
\vspace{\baselineskip}
\vspace{\baselineskip}
\vspace{\baselineskip}
\vspace{\baselineskip}
\begin{justify}
{\fontsize{11pt}{13.2pt}\selectfont \textbf{Table 2.} Overview of AI applications in ultrasound elastography classification.\par}
\end{justify}\par

%%%%%%%%%%%%%%%%%%%% Table No: 2 starts here %%%%%%%%%%%%%%%%%%%%
\begin{table}[H]
 			\centering
\begin{tabular}{p{0.5in}p{0.66in}p{1.92in}p{0.55in}p{1.8in}}
\hline
%row no:1
\multicolumn{1}{|p{0.5in}}{{\fontsize{10pt}{12.0pt}\selectfont \textbf{Ref}}} & 
\multicolumn{1}{|p{0.66in}}{{\fontsize{10pt}{12.0pt}\selectfont \textbf{Year}}} & 
\multicolumn{1}{|p{1.92in}}{{\fontsize{10pt}{12.0pt}\selectfont \textbf{Network}}} & 
\multicolumn{1}{|p{0.55in}}{{\fontsize{10pt}{12.0pt}\selectfont \textbf{ROI}}} & 
\multicolumn{1}{|p{1.8in}|}{{\fontsize{10pt}{12.0pt}\selectfont \textbf{$\#$  of samples in training/testing datasets}}} \\
\hhline{-----}
%row no:2
\multicolumn{1}{|p{0.5in}}{{\fontsize{10pt}{12.0pt}\selectfont [85]}} & 
\multicolumn{1}{|p{0.66in}}{{\fontsize{10pt}{12.0pt}\selectfont 2017}} & 
\multicolumn{1}{|p{1.92in}}{{\fontsize{10pt}{12.0pt}\selectfont SVM, Naïve Bayes, Random Forest and K-Nearest}} & 
\multicolumn{1}{|p{0.55in}}{{\fontsize{10pt}{12.0pt}\selectfont Liver }} & 
\multicolumn{1}{|p{1.8in}|}{{\fontsize{10pt}{12.0pt}\selectfont 513 subjects} \par {\fontsize{10pt}{12.0pt}\selectfont  }} \\
\hhline{-----}
%row no:3
\multicolumn{1}{|p{0.5in}}{{\fontsize{10pt}{12.0pt}\selectfont [86]}} & 
\multicolumn{1}{|p{0.66in}}{{\fontsize{10pt}{12.0pt}\selectfont 2017}} & 
\multicolumn{1}{|p{1.92in}}{{\fontsize{10pt}{12.0pt}\selectfont SVM}} & 
\multicolumn{1}{|p{0.55in}}{{\fontsize{10pt}{12.0pt}\selectfont Liver}} & 
\multicolumn{1}{|p{1.8in}|}{{\fontsize{10pt}{12.0pt}\selectfont 126 subjects}} \\
\hhline{-----}
%row no:4
\multicolumn{1}{|p{0.5in}}{{\fontsize{10pt}{12.0pt}\selectfont [87]}} & 
\multicolumn{1}{|p{0.66in}}{{\fontsize{10pt}{12.0pt}\selectfont 2012}} & 
\multicolumn{1}{|p{1.92in}}{{\fontsize{10pt}{12.0pt}\selectfont Naïve Bayes, logsitc regression}} & 
\multicolumn{1}{|p{0.55in}}{{\fontsize{10pt}{12.0pt}\selectfont Breast}} & 
\multicolumn{1}{|p{1.8in}|}{{\fontsize{10pt}{12.0pt}\selectfont leave-one-out}} \\
\hhline{-----}
%row no:5
\multicolumn{1}{|p{0.5in}}{{\fontsize{10pt}{12.0pt}\selectfont [88]}} & 
\multicolumn{1}{|p{0.66in}}{{\fontsize{10pt}{12.0pt}\selectfont 2019}} & 
\multicolumn{1}{|p{1.92in}}{{\fontsize{10pt}{12.0pt}\selectfont GoogLeNet}} & 
\multicolumn{1}{|p{0.55in}}{{\fontsize{10pt}{12.0pt}\selectfont Liver }} & 
\multicolumn{1}{|p{1.8in}|}{{\fontsize{10pt}{12.0pt}\selectfont 200 subjects}} \\
\hhline{-----}
%row no:6
\multicolumn{1}{|p{0.5in}}{{\fontsize{10pt}{12.0pt}\selectfont [89-91]}} & 
\multicolumn{1}{|p{0.66in}}{{\fontsize{10pt}{12.0pt}\selectfont 2016, 2018, 2019}} & 
\multicolumn{1}{|p{1.92in}}{{\fontsize{10pt}{12.0pt}\selectfont DPN}} & 
\multicolumn{1}{|p{0.55in}}{{\fontsize{10pt}{12.0pt}\selectfont Breast }} & 
\multicolumn{1}{|p{1.8in}|}{{\fontsize{10pt}{12.0pt}\selectfont 227 images, 5-fold CV}} \\
\hhline{-----}
%row no:7
\multicolumn{1}{|p{0.5in}}{{\fontsize{10pt}{12.0pt}\selectfont [92]}} & 
\multicolumn{1}{|p{0.66in}}{{\fontsize{10pt}{12.0pt}\selectfont 2016}} & 
\multicolumn{1}{|p{1.92in}}{{\fontsize{10pt}{12.0pt}\selectfont S-DPN}} & 
\multicolumn{1}{|p{0.55in}}{{\fontsize{10pt}{12.0pt}\selectfont Prostate}} & 
\multicolumn{1}{|p{1.8in}|}{{\fontsize{10pt}{12.0pt}\selectfont N/A}} \\
\hhline{-----}
%row no:8
\multicolumn{1}{|p{0.5in}}{{\fontsize{10pt}{12.0pt}\selectfont [93]}} & 
\multicolumn{1}{|p{0.66in}}{{\fontsize{10pt}{12.0pt}\selectfont 2016}} & 
\multicolumn{1}{|p{1.92in}}{{\fontsize{10pt}{12.0pt}\selectfont VGG}} & 
\multicolumn{1}{|p{0.55in}}{{\fontsize{10pt}{12.0pt}\selectfont Foot}} & 
\multicolumn{1}{|p{1.8in}|}{{\fontsize{10pt}{12.0pt}\selectfont 342 images, 90/10, 3-fold CV}} \\
\hhline{-----}
%row no:9
\multicolumn{1}{|p{0.5in}}{{\fontsize{10pt}{12.0pt}\selectfont [94, 95]}} & 
\multicolumn{1}{|p{0.66in}}{{\fontsize{10pt}{12.0pt}\selectfont 2016, 2020}} & 
\multicolumn{1}{|p{1.92in}}{{\fontsize{10pt}{12.0pt}\selectfont DNN}} & 
\multicolumn{1}{|p{0.55in}}{{\fontsize{10pt}{12.0pt}\selectfont Lung}} & 
\multicolumn{1}{|p{1.8in}|}{{\fontsize{10pt}{12.0pt}\selectfont 77 subjects, 5-fold CV}} \\
\hhline{-----}

\end{tabular}
 \end{table}

%%%%%%%%%%%%%%%%%%%% Table No: 2 ends here %%%%%%%%%%%%%%%%%%%%

\vspace{\baselineskip}

%\end{enumerate}
%\end{enumerate}
%\end{enumerate}
%\end{enumerate}

\paragraph{Spectral-based parameterization and envelope statistics}
\begin{justify}
{\fontsize{11pt}{13.2pt}\selectfont The Nakagami parametric map obtained based on envelope statistics illustrates the tissue microstructure scattering properties. Table 3 shows a list of selected references that used AI in spectral-based parameterization and envelope statistics. Destrempes \textit{et al}. used thirteen features in terms of statistical and spectral backscatter properties of tissues, maximum elasticity and total attenuation coefficient to classify solid breast lesions using a random forest model [96]. Via a feature selection technique, BI-RADS category of elasticity and QUS features were used as input features. This technique obtained an AUC of 0.97 with sensitivity of 98$\%$  and specificity of 75.9$\%$ . Cardinal \textit{et al}. used the homodyned-K parametric maps, Nakagami parametric maps and log-compressed B-mode images to compute QUS features and classify components of carotid plaques via a random forest model [97]. Feature selection technique was performed to reduce the number of QUS parameters to a maximum of three per classification task. The obtained AUCs were 0.79 (lipid area $ \geq $  10$\%$ ), 0.9 (lipid area $ \geq $  25$\%$ ), 0.95 (calcium area), 0.97 (ruptured fibrous cap) and 0.91 (plaque) based on ultrasound elastography, homodyned-K parameters and echogenicity. Gangeh \textit{et al}. proposed a computer-aided-prognosis system for the early prediction of cell death levels based on QUS spectroscopy [98]. Spectral parameters such as midband fit and spectral intercept were calculated to form parametric maps. Local binary patterns (LBPs) were used for texture analysis on parametric maps given its advantage of unifying statistical and structural approaches. A kernel-based metric called maximum mean discrepancy (MMD) was adopted for assessing the treatment effectiveness. Naïve Bayes classifier was used for categorization of cell death and a linear support vector regressor was used for prediction of continuous cell death levels. The combinations of histogram of intensity (HistInt) and LBPs as features with the MMD as the dissimilarity measure achieved the best performance in classification of cell death levels at 20$\%$  (accuracy of 85$\%$  and AUC of 0.873) and at a 40$\%$  threshold (accuracy of 85.2$\%$  and AUC of 0.869). The highest correlation coefficient (r = 0.681) in cell death prediction based on this combination was also obtained. Ghorayeb \textit{et al}. developed a platform known as $``$quantitative ultrasound fetal lung maturity analysis$"$ , which combines various image texture extractors based on fetal lung ultrasound images and ML algorithms, to predict neonatal respiratory morbidity such as respiratory distress syndrome or transient tachypnea for the newborn [99, 100].\ The final classification algorithm was a sequence of combining various ML algorithms and achieved a sensitivity of 86.2$\%$  and a specificity of 87.0$\%$ .  Caxinha \textit{et al}. used the Naïve Bayes, K-Nearest Neighbor (KNN), SVM, and Fisher Linear Discriminant (FLD) classifiers for cataract classification based on ninety-seven parameters extracted from acoustical, spectral and image textural analyses [101]. A textural analysis was performed to extract features from the B-scan images while Nakagami distribution was used to estimate the probability density function of the backscattered signals. In the classification between healthy and cataractous lenses, all four classifiers showed good performances with F-measure over 92.68$\%$  with KNN obtained the highest sensitivity of 95.71$\%$  and specificity of 85.55$\%$ . In the classification between initial and severe cataractous lenses, only SVM showed a F-measure of 90.62$\%$  with sensitivity of 98.3$\%$  and specificity of 80.59$\%$ . \par}
\end{justify}\par

\begin{justify}
{\fontsize{11pt}{13.2pt}\selectfont Tadayyon \textit{et al}. predicted the breast tumor response to chemotherapy using a ANN model based on QUS measurements, such as midband fit (MBF), spectral slope (SS), spectral intercept (SI), spacing among scatters (SAS), attenuation coefficient estimate (ACE), average scatterer diameter (ASD), average acoustic concentration (AAC) and texture features [102]. The accuracy (96 $ \pm $  6$\%$ ) and AUC (0.96 $ \pm $  0.08) using the ANN were higher than those using the KNN model (accuracy: 65 $ \pm $  10$\%$ , AUC: 0.67 $ \pm $  0.14). Feleppa \textit{et al}. used the spectral intercept, spectral midband, and prostate-specific antigen levels for classifying prostate cancer with four nonlinear classifiers (ANN, logitboost algorithms (LBA), SVM, and RBM) [103]. The obtained AUCs were 0.84 $ \pm $  0.02, 0.87 $ \pm $  0.04, 0.89 $ \pm $  0.04, and 0.91 $ \pm $  0.04 for the ANN, LBA, SVM and RBM, respectively. Byra \textit{et al}. adopted a CNN model trained on the Nakagami parametric map for breast lesion classification [104].\ This model achieved an AUC of 0.912 $ \pm $  0.005 with a sensitivity of 82.4$\%$  and a specificity of 83.3$\%$ .  Lekadir \textit{et al}. utilized a CNN model for identifying lipid core, fibrous cap and calcified tissue regions based on carotid ultrasound images [105]. Via a patch-based technique, the sample was translated into 90,000 input imaging patches for the CNN model. The obtained correlation coefficients were 0.92 for the lipid core (sensitivity of 0.83 and specificity of 0.9), 0.87 for the fibrous cap (sensitivity of 0.7 and specificity of 0.8), 0.93 for the calcified tissue (sensitivity of 0.83 and specificity of 0.9). \par}
\end{justify}\par

%\subsubsection*{Table 3. Overview of AI applications in spectral-based parameterization and envelope statistics.}
%\addcontentsline{toc}{subsubsection}{Table 3. Overview of AI applications in spectral-based parameterization and envelope statistics.}
\vspace{\baselineskip}
\vspace{\baselineskip}
\vspace{\baselineskip}
\vspace{\baselineskip}
\vspace{\baselineskip}
\vspace{\baselineskip}
\vspace{\baselineskip}
\vspace{\baselineskip}
\vspace{\baselineskip}
\vspace{\baselineskip}
\vspace{\baselineskip}
\vspace{\baselineskip}
\vspace{\baselineskip}
\vspace{\baselineskip}
\vspace{\baselineskip}
\begin{justify}
{\fontsize{11pt}{13.2pt}\selectfont \textbf{Table 3.} Overview of AI applications in spectral-based parameterization and envelope statistics.\par}
\end{justify}\par

%%%%%%%%%%%%%%%%%%%% Table No: 3 starts here %%%%%%%%%%%%%%%%%%%%

\begin{table}[H]
 			\centering
\begin{tabular}{p{0.49in}p{0.3in}p{0.99in}p{0.97in}p{2.63in}}
\hline
%row no:1
\multicolumn{1}{|p{0.49in}}{{\fontsize{10pt}{12.0pt}\selectfont \textbf{Ref}}} & 
\multicolumn{1}{|p{0.3in}}{{\fontsize{10pt}{12.0pt}\selectfont \textbf{Year}}} & 
\multicolumn{1}{|p{0.99in}}{{\fontsize{10pt}{12.0pt}\selectfont \textbf{Network}}} & 
\multicolumn{1}{|p{0.97in}}{{\fontsize{10pt}{12.0pt}\selectfont \textbf{ROI}}} & 
\multicolumn{1}{|p{2.63in}|}{{\fontsize{10pt}{12.0pt}\selectfont \textbf{$\#$  of samples in training/testing datasets}}} \\
\hhline{-----}
%row no:2
\multicolumn{1}{|p{0.49in}}{{\fontsize{10pt}{12.0pt}\selectfont [96]}} & 
\multicolumn{1}{|p{0.3in}}{{\fontsize{10pt}{12.0pt}\selectfont 2020}} & 
\multicolumn{1}{|p{0.99in}}{{\fontsize{10pt}{12.0pt}\selectfont Random forest}} & 
\multicolumn{1}{|p{0.97in}}{{\fontsize{10pt}{12.0pt}\selectfont Breast}} & 
\multicolumn{1}{|p{2.63in}|}{{\fontsize{10pt}{12.0pt}\selectfont 103 subjects, bootstrap CV}} \\
\hhline{-----}
%row no:3
\multicolumn{1}{|p{0.49in}}{{\fontsize{10pt}{12.0pt}\selectfont [97]}} & 
\multicolumn{1}{|p{0.3in}}{{\fontsize{10pt}{12.0pt}\selectfont 2018}} & 
\multicolumn{1}{|p{0.99in}}{{\fontsize{10pt}{12.0pt}\selectfont Random forest}} & 
\multicolumn{1}{|p{0.97in}}{{\fontsize{10pt}{12.0pt}\selectfont Carotid artery}} & 
\multicolumn{1}{|p{2.63in}|}{{\fontsize{10pt}{12.0pt}\selectfont 66 subjects, bootstrap CV, }} \\
\hhline{-----}
%row no:4
\multicolumn{1}{|p{0.49in}}{{\fontsize{10pt}{12.0pt}\selectfont [98]}} & 
\multicolumn{1}{|p{0.3in}}{{\fontsize{10pt}{12.0pt}\selectfont 2016}} & 
\multicolumn{1}{|p{0.99in}}{{\fontsize{10pt}{12.0pt}\selectfont Naïve Bayes, SVR}} & 
\multicolumn{1}{|p{0.97in}}{{\fontsize{10pt}{12.0pt}\selectfont Cell }} & 
\multicolumn{1}{|p{2.63in}|}{{\fontsize{10pt}{12.0pt}\selectfont N/A, Leave-one-out CV}} \\
\hhline{-----}
%row no:5
\multicolumn{1}{|p{0.49in}}{{\fontsize{10pt}{12.0pt}\selectfont [99, 100]}} & 
\multicolumn{1}{|p{0.3in}}{{\fontsize{10pt}{12.0pt}\selectfont 2017, 2015}} & 
\multicolumn{1}{|p{0.99in}}{{\fontsize{10pt}{12.0pt}\selectfont SVR, SVM}} & 
\multicolumn{1}{|p{0.97in}}{{\fontsize{10pt}{12.0pt}\selectfont Respiratory}} & 
\multicolumn{1}{|p{2.63in}|}{{\fontsize{10pt}{12.0pt}\selectfont 425 subjects, 144 images}} \\
\hhline{-----}
%row no:6
\multicolumn{1}{|p{0.49in}}{{\fontsize{10pt}{12.0pt}\selectfont [102]}} & 
\multicolumn{1}{|p{0.3in}}{{\fontsize{10pt}{12.0pt}\selectfont 2019}} & 
\multicolumn{1}{|p{0.99in}}{{\fontsize{10pt}{12.0pt}\selectfont ANN}} & 
\multicolumn{1}{|p{0.97in}}{{\fontsize{10pt}{12.0pt}\selectfont Breast}} & 
\multicolumn{1}{|p{2.63in}|}{{\fontsize{10pt}{12.0pt}\selectfont 100 subjects, 70/15/15}} \\
\hhline{-----}
%row no:7
\multicolumn{1}{|p{0.49in}}{{\fontsize{10pt}{12.0pt}\selectfont [103]}} & 
\multicolumn{1}{|p{0.3in}}{{\fontsize{10pt}{12.0pt}\selectfont 2019}} & 
\multicolumn{1}{|p{0.99in}}{{\fontsize{10pt}{12.0pt}\selectfont ANN, logitboost algorithms, SVM, and RBM}} & 
\multicolumn{1}{|p{0.97in}}{{\fontsize{10pt}{12.0pt}\selectfont Prostate}} & 
\multicolumn{1}{|p{2.63in}|}{{\fontsize{10pt}{12.0pt}\selectfont 64 subjects}} \\
\hhline{-----}
%row no:8
\multicolumn{1}{|p{0.49in}}{{\fontsize{10pt}{12.0pt}\selectfont [104]}} & 
\multicolumn{1}{|p{0.3in}}{{\fontsize{10pt}{12.0pt}\selectfont 2017}} & 
\multicolumn{1}{|p{0.99in}}{{\fontsize{10pt}{12.0pt}\selectfont CNN}} & 
\multicolumn{1}{|p{0.97in}}{{\fontsize{10pt}{12.0pt}\selectfont Breast}} & 
\multicolumn{1}{|p{2.63in}|}{{\fontsize{10pt}{12.0pt}\selectfont 5-fold CV}} \\
\hhline{-----}
%row no:9
\multicolumn{1}{|p{0.49in}}{{\fontsize{10pt}{12.0pt}\selectfont [105]}} & 
\multicolumn{1}{|p{0.3in}}{{\fontsize{10pt}{12.0pt}\selectfont 2016}} & 
\multicolumn{1}{|p{0.99in}}{{\fontsize{10pt}{12.0pt}\selectfont CNN}} & 
\multicolumn{1}{|p{0.97in}}{{\fontsize{10pt}{12.0pt}\selectfont Carotid artery}} & 
\multicolumn{1}{|p{2.63in}|}{{\fontsize{10pt}{12.0pt}\selectfont 56 subjects, 5-fold CV}} \\
\hhline{-----}

\end{tabular}
 \end{table}

%%%%%%%%%%%%%%%%%%%% Table No: 3 ends here %%%%%%%%%%%%%%%%%%%%

\vspace{\baselineskip}
\subsubsection{Radiomics}
\begin{justify}
{\fontsize{11pt}{13.2pt}\selectfont Zhou \textit{et al}. used a CNN-based radiomics approach on SWE images to diagnose breast cancer without segmentation [90]. It also compared the performances with different inputs, such as superposed image, pure SWE data in RGB and the recoded SWE data in HUE. An accuracy of 95.8$\%$ , a sensitivity of 96.2$\%$  and a specificity of 95.7$\%$  were obtained in the test dataset.\  Axillary lymph node (ALN) metastasis is important in guiding treatment of breast cancer. Sun \textit{et al}.  compared the performances of CNN and radiomics approach in predicting ALN metastasis using breast ultrasound [106]. The CNN model was built in DenseNet while radiomics models were built in random forest models. The obtained results from CNNs showed numerically better overall performance compared with radiomics models in predicting ALN metastasis in breast cancer. Based on SWE images, Wang \textit{et al}. used a DL radiomics of elastography for staging liver fibrosis in chronic hepatitis B [48]. The radiomics strategy for quantitative analysis of the heterogeneity in 2D SWE images was adopted and the obtained results showed with AUCs of 0.97 for F4 (95$\%$  confidence interval (CI) 0.94 to 0.99), 0.98 for $ \geq $  F3 (95$\%$  CI 0.96 to 1.00) and 0.85 for $ \geq $  F2 (95$\%$  CI 0.81 to 0.89), indicating its better performance than liver stiffness measurement.\par}
\end{justify}\par

\subsubsection{Super resolution}
\begin{justify}
{\fontsize{11pt}{13.2pt}\selectfont In prostate brachytherapy, US imaging is routinely used to visualize and guide seed or needle placement. Due to limited acquisition time and hardware limitation, a routine 3D US with 2-5mm slice thickness is captured. 3D image with low resolution will degrade the accuracy of prostate contouring and needle/seed detection in prostate brachytherapy. Lei \textit{et al}. developed a DL-based method to reconstruct high-resolution images from routinely captured prostate US images during brachytherapy [107, 108]. They used deep attentional GAN model trained on the patched extracted from the low-resolution US images to generate patches which were then fused into high-resolution US images. The results showed that a mean absolute error (MAE) and peak signal-to-noise ratio (PSNR) of image intensity profiles between original and reconstructed images were 6.5 $ \pm $  0.5 and 38.0 $ \pm $  2.4 dB. Van Sloun \textit{et al}. proposed to use a CNN to map low-resolution contrast-enhanced ultrasound image to highly resolvable localizations [109, 110]. The architecture of this network is similar to U-net, in which there are encoder compressing the input frames into a latent representation and decoder decoding this representation into a high-resolution output. The input image dimensions were scaled up eight times via this network. Moreover, the noise was suppressed using this network in the compact latent space. \par}
\end{justify}\par

\subsubsection{Segmentation}
\begin{justify}
{\fontsize{11pt}{13.2pt}\selectfont Segmentation using AI provides an objective and reproducible technique for localizing the region of interest, significantly reducing the time and alleviating physicians’ burden. It can be used in motion tracking in ultrasound B-mode, echocardiography, etc. \par}
\end{justify}\par

%\begin{enumerate}[label*={\fontsize{11pt}{11pt}\selectfont \arabic*.}]
	%\item \begin{enumerate}[label*={\fontsize{11pt}{11pt}\selectfont \arabic*.}]
	%\item \begin{enumerate}[label*={\fontsize{11pt}{11pt}\selectfont \arabic*.}]
	%\item \begin{enumerate}[label*={\fontsize{11pt}{11pt}\selectfont \arabic*.}]
	%\item {\fontsize{11pt}{13.2pt}\selectfont B-mode motion tracking\par}\par

\paragraph{B-mode motion tracking}
\vspace{\baselineskip}
\begin{justify}
{\fontsize{11pt}{13.2pt}\selectfont Table 4 shows a list of selected references that used AI in B-mode motion tracking. Rangamani \textit{et al}. employed a CNN coupled with RNN for landmark detection and target tracking [111]. The CNN encoder-decoder was used for landmark detecton while RNN was used for encoding information from previous video frames to track objects. Gomariz \textit{et al}. proposed a fully-convolutional Siamese network to localize and track anatomical targets [112].\ A temporal consistency model was employed as a location prior and combined with the network-predicted location probability map for tracking the target in the ultrasound sequences.  Huang \textit{et al}. developed a DL-based real-time motion tracking technique for ultrasound image-guided radiation therapy by combining the attention-aware fully CNN (FCNN) and the convolutional long short-term memory network (CLSTM) [113, 114]. FCNN extracted spatial features while CLSTM refined the features via computing the saliency mask. In order to improve the training convergence and alleviate over/underfitting, multitask losses such as bounding box loss, localization loss, saliency loss and adaptive loss weighting term was employed. \par}
\end{justify}\par

\begin{justify}
{\fontsize{11pt}{13.2pt}\selectfont Prostate cancer continues to be the second leading cause of cancer death in men [115]. Due to its inexpensiveness and easiness to use, ultrasound imaging is the main imaging modality for prostate cancer diagnosis and treatment. Accurate segmentation of the prostate is important to biopsy needle placement and brachytherapy treatment planning. For example, segmentation of prostate will be beneficial for clinicians to quantify the volume of the prostate gland so as to plan treatment in High-Dose-Rate (HDR) and Low-Dose-Rate (LDR) brachytherapy. Manual segmentation during biopsy or radiation therapy can be time-consuming. Transrectal US (TRUS) is the standard imaging modality for the image-guided prostate-cancer interventions due to its versatility and real-time capability. Ghose \textit{et al}. proposed a supervised learning scheme via a random forest model for automatic initialization and propagation of statistical shape and appearance model [116-118]. The appearance model was first derived from the posterior probabilities obtained from the random forest classifier. This probabilistic information was then used for the initialization and propagation of the statistical model. This method achieved a mean Dice Similarity Coefficient (DSC) value of 0.96 $ \pm $  0.01 on 24 images with considerable shape, size and intensity variations obtained from TRUS. The same group later proposed an approach of building multiple mean parametric models derived from principal component analysis of shape and posterior probabilities in a multi-resolution framework [119, 120]. The model parameters were then modified with the prior knowledge of the optimization space to achieve optimal prostate segmentation. Moreover, multiple mean models derived from spectral clustering of combined shape and appearance parameters were used in parallel to improve segmentation accuracies. Sahba \textit{et al}. proposed a reinforcement learning (RL) framework for prostate segmentation in 2D TRUS slices [121]. This algorithm was used to find the appropriate local values for sub-images based on reward/punishment. The obtained result was assessed in terms of area overlap and was 90.96 $ \pm $  2.2$\%$ .\par}
\end{justify}\par

\begin{justify}
{\fontsize{11pt}{13.2pt}\selectfont Ghavami \textit{et al}. used a CNN for automatic prostate segmentation in 2D TRUS slices and 3D TRUS volumes [122, 123]. They designed the network to be able to incorporate 3D spatial information by taking one or more TRUS slices neighboring each slice to be segmented as input. This method achieved a mean 2D DSC value of 0.91 $ \pm $  0.12 and a mean absolute boundary segmentation error of 1.23 $ \pm $  1.46 mm. Lei \textit{et al}. developed a 3D patch-based V-Net for prostate segmentation [124]. A 3D supervision mechanism was integrated into the V-Net stages to deal with the optimization difficulties when training a deep network with limited training data. The patches were extracted from the ultrasound image as the input. A binary cross-entropy loss and a batch-based Dice loss were combined into the stage-wise hybrid loss function for deep supervision learning. The segmented prostate volume was reconstructed using patch fusion and refined via a contour refinement processing. They validated the accuracy of this technique via comparing to manual segmentation and demonstrated its clinical feasibility. Karimi \textit{et al}. adopted an adaptive sampling strategy in training CNN to pay more attention to images that are difficult to segment [125]. Moreover, they trained a CNN ensemble and the disagreement among the ensemble was used to identify uncertain segmentation and to estimate a segmentation uncertainty map. Wang \textit{et al}. developed a 3D deep neural network coupled with attention modules for prostate segmentation in TRUS by fully exploiting the complementary information encoded in different layers of the CNN [126, 127]. The attention module used the attention mechanism to selectively leverage the multilevel features integrated from different layers to refine the features at each individual layer, suppressing the noise at shallow layers and increasing more prostate details into features at deep layers. The efficacy of the network was evaluated on challenging prostate TRUS images and they demonstrated that this network outperformed state-of-the-art methods by a large margin. \par}
\end{justify}\par

\begin{justify}
{\fontsize{11pt}{13.2pt}\selectfont In contrast to CNN, Hassanien \textit{et al}. adopted a pulse-coupled neural networks (PCNNs) for accurate boundary detection on the prostate ultrasound images [128]. In order to increase the contrast of the ultrasound image, the intensity values of the images were firstly adjusted via the PCNN with median filter. Then, PCNN was used for prostate segmentation. PCNN sensitivity to the setting of the PCNN parameters was eliminated by combining adjustment and segmentation with PCNN. One issue with prostate segmentation in ultrasound images is boundary incompleteness. In order to solve this issue, Yang \textit{et al}. proposed a fine-grained boundary completion recurrent neural network (BCRNN) [129, 130]. This framework integrated feature extraction and shape prior to explore and estimate the complete boundary in a sequential manner. The static prostate ultrasound images were serialized into dynamic sequence and shape priors were sequentially explored to infer prostate shapes. A multi-view fusion strategy was adopted to merge shape predictions obtained from different perspectives into a comprehensive prediction. Moreover, a multi-scale Auto-Context scheme was implemented to gain incremental refinement on prostate shape predictions. Anas \textit{et al}. combined CNN and RNN to extract both spatial and temporal features from a series of TRUS images [131]. Residual convolution in the RNN was used to improve optimization. Both dense and sparse sampling of the input ultrasound sequence were performed to assess the robustness of the network to ultrasound artifacts.\par}
\end{justify}\par

\begin{justify}
{\fontsize{11pt}{13.2pt}\selectfont Breast cancer is one of the leading causes of cancer death among women worldwide. In clinical routine, automatic breast ultrasound image segmentation is very challenging and essential for cancer diagnosis and treatment planning. Braz \textit{et al}. used the unsupervised K-means and the supervised SVM models to classify the image pixel descriptors for breast gland segmentation on breast ultrasound (BUS) images [132]. Image preprocessing techniques were performed to generate the image pixel descriptors. Jiang \textit{et al}. formulated the breast tumor detection as a two-step learning problem [133]. They first used the Adaboost classifier on Harr-like features to detect a preliminary set of tumor regions. The preliminary tumor regions were screened with a SVM classifier based on the quantized intensity features. They finally used the random walks segmentation algorithm to retrieve the boundary of tumor region. The obtained results demonstrated that the proposed algorithm was able to localize and segment the tumor regions with high accuracy. Torbati \textit{et al}. modified a self-organizing network to segment BUS images [134, 135]. A 2D discrete wavelet transform was used to build the input feature space of the network. Huang \textit{et al}. proposed an automatic interaction scheme to segment the lesions in BUS images [136]. The ultrasound image was firstly filtered with a total-variation model to reduce speckle noise. Then a robust graph-based segmentation method was used to segment the image with an object recognition technique to identify the breast tumor region. Finally, an active contour model was used to refine the tumor boundary. Daoud \textit{et al}. proposed an algorithm for segmenting lesions in BUS images by combining image boundary and region information [137]. This framework decomposed the ultrasound image into a set of superpixels using the Normalized Cuts method. Then a SVM classifier was used to estimate the tumor likelihood of each superpixel\ based on texture features.  \par}
\end{justify}\par

\begin{justify}
{\fontsize{11pt}{13.2pt}\selectfont CNN has been used for breast segmentation and distinguishing functional tissues on 3D ultrasound images. Lei \textit{et al}. proposed a boundary regularized deep convolutional encoder-decoder network (ConvEDNet) for the segmentation of breast anatomical layers in the automated whole breast ultrasound (AWBUS) images [138]. The training of ConvEDNet was regularized by the geometric constraints with the deep supervision technique for better withstand of intrinsic speckle noise and posterior acoustic shadows in ultrasound images and further boosted with the adaptive domain transfer (ADT). The ADT is a two-stage domain transfer technique for better landing the encoder on the ultrasound domain. Singh \textit{et al}. proposed to add a dilated convolutional layer to the conditional GAN segmentation model for extracting tumor features at various resolutions of BUS images [139]. A channel-wise weighting block was also proposed to add in the network to automatically rebalance the impact of each of the highest-level encoded features. This model outperformed the state-of-the-art segmentation models in terms of the DSC metrics. Hu \textit{et al}. proposed to combine a dilated fully convolutional network (DFCN) with a phase-based active contour model for automatic tumor segmentation in BUS images [140]. The DFCN is an improved fully CNN with dilated convolution in deeper layers, requiring few parameters and obtaining a larger receptive field. Yap \textit{et al}. used the fully convolutional networks (FCNs) for semantic segmentation of breast lesions on BUS images [141]. They used pretrained models based on ImageNet on two datasets consisted of both malignant and benign lesions. To overcome the issues of large, weakly annotated datasets or overfitting with small, strongly annotated datasets, Shin \textit{et al}. proposed a systematic weakly and semi-supervised training scenario on a weakly annotated dataset together with a smaller strongly annotated dataset in a hybrid manner [142]. The results trained with only 10 strongly annotated images along with weakly annotated images were comparable to results trained from 800 strongly annotated images, with the 95$\%$  confidence interval of difference -3$\%$  - 5$\%$  in terms of the correct localization measure. \par}
\end{justify}\par

%\end{enumerate}
%\end{enumerate}
%\end{enumerate}
%\end{enumerate}\subsubsection*{Table 4. Overview of AI applications in B-mode motion tracking.}
%\addcontentsline{toc}{subsubsection}{Table 4. Overview of AI applications in B-mode motion tracking.}

\begin{justify}
{\fontsize{11pt}{13.2pt}\selectfont \textbf{Table 4.} Overview of AI applications in B-mode motion tracking.\par}
\end{justify}\par

%%%%%%%%%%%%%%%%%%%% Table No: 4 starts here %%%%%%%%%%%%%%%%%%%%

{
\setlength\extrarowheight{3pt}
\begin{longtable}{p{0.42in}p{0.68in}p{1.36in}p{0.52in}p{2.68in}}

\endfirsthead
%\multicolumn{5}{c}{\textit{continued from previous page}}\hhline
%\endhead
%\multicolumn{5}{r}{\textit{continued on next page}} \\
%\endfoot
%\endlastfoot\hline
%row no:1
\hline
\multicolumn{1}{|p{0.42in}}{{\fontsize{10pt}{12.0pt}\selectfont \textbf{Ref}}} & 
\multicolumn{1}{|p{0.68in}}{{\fontsize{10pt}{12.0pt}\selectfont \textbf{Year}}} & 
\multicolumn{1}{|p{1.36in}}{{\fontsize{10pt}{12.0pt}\selectfont \textbf{Network}}} & 
\multicolumn{1}{|p{0.52in}}{{\fontsize{10pt}{12.0pt}\selectfont \textbf{ROI}}} & 
\multicolumn{1}{|p{2.68in}|}{{\fontsize{10pt}{12.0pt}\selectfont \textbf{$\#$  of samples in training/testing datasets}}} \\
\hhline{-----}
%row no:2
\multicolumn{1}{|p{0.42in}}{{\fontsize{10pt}{12.0pt}\selectfont [112]}} & 
\multicolumn{1}{|p{0.68in}}{{\fontsize{10pt}{12.0pt}\selectfont 2019}} & 
\multicolumn{1}{|p{1.36in}}{{\fontsize{10pt}{12.0pt}\selectfont FCN}} & 
\multicolumn{1}{|p{0.52in}}{{\fontsize{10pt}{12.0pt}\selectfont Liver }} & 
\multicolumn{1}{|p{2.68in}|}{{\fontsize{10pt}{12.0pt}\selectfont 24/40 sequences}} \\
\hhline{-----}
%row no:3
\multicolumn{1}{|p{0.42in}}{{\fontsize{10pt}{12.0pt}\selectfont [113, 114]}} & 
\multicolumn{1}{|p{0.68in}}{{\fontsize{10pt}{12.0pt}\selectfont 2019}} & 
\multicolumn{1}{|p{1.36in}}{{\fontsize{10pt}{12.0pt}\selectfont FCNN}} & 
\multicolumn{1}{|p{0.52in}}{{\fontsize{10pt}{12.0pt}\selectfont Liver}} & 
\multicolumn{1}{|p{2.68in}|}{{\fontsize{10pt}{12.0pt}\selectfont 25/39 sequences}} \\
\hhline{-----}
%row no:4
\multicolumn{1}{|p{0.42in}}{{\fontsize{10pt}{12.0pt}\selectfont [116-118]}} & 
\multicolumn{1}{|p{0.68in}}{{\fontsize{10pt}{12.0pt}\selectfont 2012}} & 
\multicolumn{1}{|p{1.36in}}{{\fontsize{10pt}{12.0pt}\selectfont Random forest}} & 
\multicolumn{1}{|p{0.52in}}{{\fontsize{10pt}{12.0pt}\selectfont Prostate}} & 
\multicolumn{1}{|p{2.68in}|}{{\fontsize{10pt}{12.0pt}\selectfont 24 images, leave-one-out}} \\
\hhline{-----}
%row no:5
\multicolumn{1}{|p{0.42in}}{{\fontsize{10pt}{12.0pt}\selectfont [119, 120]}} & 
\multicolumn{1}{|p{0.68in}}{{\fontsize{10pt}{12.0pt}\selectfont 2013, 2011}} & 
\multicolumn{1}{|p{1.36in}}{{\fontsize{10pt}{12.0pt}\selectfont Random forest}} & 
\multicolumn{1}{|p{0.52in}}{{\fontsize{10pt}{12.0pt}\selectfont Prostate}} & 
\multicolumn{1}{|p{2.68in}|}{{\fontsize{10pt}{12.0pt}\selectfont 23 datasets, leave-one-patient-out}} \\
\hhline{-----}
%row no:6
\multicolumn{1}{|p{0.42in}}{{\fontsize{10pt}{12.0pt}\selectfont [121]}} & 
\multicolumn{1}{|p{0.68in}}{{\fontsize{10pt}{12.0pt}\selectfont 2008}} & 
\multicolumn{1}{|p{1.36in}}{{\fontsize{10pt}{12.0pt}\selectfont Reinforcement learning}} & 
\multicolumn{1}{|p{0.52in}}{{\fontsize{10pt}{12.0pt}\selectfont Prostate}} & 
\multicolumn{1}{|p{2.68in}|}{{\fontsize{10pt}{12.0pt}\selectfont 8/52 images}} \\
\hhline{-----}
%row no:7
\multicolumn{1}{|p{0.42in}}{{\fontsize{10pt}{12.0pt}\selectfont [122, 123]}} & 
\multicolumn{1}{|p{0.68in}}{{\fontsize{10pt}{12.0pt}\selectfont 2018}} & 
\multicolumn{1}{|p{1.36in}}{{\fontsize{10pt}{12.0pt}\selectfont CNN}} & 
\multicolumn{1}{|p{0.52in}}{{\fontsize{10pt}{12.0pt}\selectfont Prostate}} & 
\multicolumn{1}{|p{2.68in}|}{{\fontsize{10pt}{12.0pt}\selectfont 110 subjects, 10-fold CV}} \\
\hhline{-----}
%row no:8
\multicolumn{1}{|p{0.42in}}{{\fontsize{10pt}{12.0pt}\selectfont [124]}} & 
\multicolumn{1}{|p{0.68in}}{{\fontsize{10pt}{12.0pt}\selectfont 2019}} & 
\multicolumn{1}{|p{1.36in}}{{\fontsize{10pt}{12.0pt}\selectfont 3D V-net}} & 
\multicolumn{1}{|p{0.52in}}{{\fontsize{10pt}{12.0pt}\selectfont Prostate}} & 
\multicolumn{1}{|p{2.68in}|}{{\fontsize{10pt}{12.0pt}\selectfont 44 subjects, 5-fold CV}} \\
\hhline{-----}
%row no:9
\multicolumn{1}{|p{0.42in}}{{\fontsize{10pt}{12.0pt}\selectfont [125]}} & 
\multicolumn{1}{|p{0.68in}}{{\fontsize{10pt}{12.0pt}\selectfont 2019}} & 
\multicolumn{1}{|p{1.36in}}{{\fontsize{10pt}{12.0pt}\selectfont CNN}} & 
\multicolumn{1}{|p{0.52in}}{{\fontsize{10pt}{12.0pt}\selectfont Prostate}} & 
\multicolumn{1}{|p{2.68in}|}{} \\
\hhline{-----}
%row no:10
\multicolumn{1}{|p{0.42in}}{{\fontsize{10pt}{12.0pt}\selectfont [126, 127]}} & 
\multicolumn{1}{|p{0.68in}}{{\fontsize{10pt}{12.0pt}\selectfont 2019,\  2018}} & 
\multicolumn{1}{|p{1.36in}}{{\fontsize{10pt}{12.0pt}\selectfont 3D DNN with attention module}} & 
\multicolumn{1}{|p{0.52in}}{{\fontsize{10pt}{12.0pt}\selectfont Prostate}} & 
\multicolumn{1}{|p{2.68in}|}{{\fontsize{10pt}{12.0pt}\selectfont 40 subjects, 4-fold CV}} \\
\hhline{-----}
%row no:11
\multicolumn{1}{|p{0.42in}}{{\fontsize{10pt}{12.0pt}\selectfont [128]}} & 
\multicolumn{1}{|p{0.68in}}{{\fontsize{10pt}{12.0pt}\selectfont 2011}} & 
\multicolumn{1}{|p{1.36in}}{{\fontsize{10pt}{12.0pt}\selectfont Pulsed CNN}} & 
\multicolumn{1}{|p{0.52in}}{{\fontsize{10pt}{12.0pt}\selectfont Prostate}} & 
\multicolumn{1}{|p{2.68in}|}{{\fontsize{10pt}{12.0pt}\selectfont 212 subjects}} \\
\hhline{-----}
%row no:12
\multicolumn{1}{|p{0.42in}}{{\fontsize{10pt}{12.0pt}\selectfont [129, 130]}} & 
\multicolumn{1}{|p{0.68in}}{{\fontsize{10pt}{12.0pt}\selectfont 2017}} & 
\multicolumn{1}{|p{1.36in}}{{\fontsize{10pt}{12.0pt}\selectfont Boundary completion RNN}} & 
\multicolumn{1}{|p{0.52in}}{{\fontsize{10pt}{12.0pt}\selectfont Prostate}} & 
\multicolumn{1}{|p{2.68in}|}{{\fontsize{10pt}{12.0pt}\selectfont 17 subjects, 2400/130 images}} \\
\hhline{-----}
%row no:13
\multicolumn{1}{|p{0.42in}}{{\fontsize{10pt}{12.0pt}\selectfont [131]}} & 
\multicolumn{1}{|p{0.68in}}{{\fontsize{10pt}{12.0pt}\selectfont 2018}} & 
\multicolumn{1}{|p{1.36in}}{{\fontsize{10pt}{12.0pt}\selectfont CNN, RNN}} & 
\multicolumn{1}{|p{0.52in}}{{\fontsize{10pt}{12.0pt}\selectfont Prostate}} & 
\multicolumn{1}{|p{2.68in}|}{{\fontsize{10pt}{12.0pt}\selectfont 2238/637/1017}} \\
\hhline{-----}
%row no:14
\multicolumn{1}{|p{0.42in}}{{\fontsize{10pt}{12.0pt}\selectfont [132]}} & 
\multicolumn{1}{|p{0.68in}}{{\fontsize{10pt}{12.0pt}\selectfont 2012}} & 
\multicolumn{1}{|p{1.36in}}{{\fontsize{10pt}{12.0pt}\selectfont K-means, SVM}} & 
\multicolumn{1}{|p{0.52in}}{{\fontsize{10pt}{12.0pt}\selectfont Breast }} & 
\multicolumn{1}{|p{2.68in}|}{{\fontsize{10pt}{12.0pt}\selectfont 61 images}} \\
\hhline{-----}
%row no:15
\multicolumn{1}{|p{0.42in}}{{\fontsize{10pt}{12.0pt}\selectfont [133]}} & 
\multicolumn{1}{|p{0.68in}}{{\fontsize{10pt}{12.0pt}\selectfont 2012}} & 
\multicolumn{1}{|p{1.36in}}{{\fontsize{10pt}{12.0pt}\selectfont Adaboost, SVM}} & 
\multicolumn{1}{|p{0.52in}}{} & 
\multicolumn{1}{|p{2.68in}|}{{\fontsize{10pt}{12.0pt}\selectfont 112 images, 4-fold CV}} \\
\hhline{-----}
%row no:16
\multicolumn{1}{|p{0.42in}}{{\fontsize{10pt}{12.0pt}\selectfont [134, 135]}} & 
\multicolumn{1}{|p{0.68in}}{{\fontsize{10pt}{12.0pt}\selectfont 2013, 2014}} & 
\multicolumn{1}{|p{1.36in}}{{\fontsize{10pt}{12.0pt}\selectfont Self-organizing network}} & 
\multicolumn{1}{|p{0.52in}}{{\fontsize{10pt}{12.0pt}\selectfont Breast}} & 
\multicolumn{1}{|p{2.68in}|}{{\fontsize{10pt}{12.0pt}\selectfont 30 BUS images}} \\
\hhline{-----}
%row no:17
\multicolumn{1}{|p{0.42in}}{{\fontsize{10pt}{12.0pt}\selectfont [136]}} & 
\multicolumn{1}{|p{0.68in}}{{\fontsize{10pt}{12.0pt}\selectfont 2015}} & 
\multicolumn{1}{|p{1.36in}}{{\fontsize{10pt}{12.0pt}\selectfont SVM}} & 
\multicolumn{1}{|p{0.52in}}{{\fontsize{10pt}{12.0pt}\selectfont Breast}} & 
\multicolumn{1}{|p{2.68in}|}{{\fontsize{10pt}{12.0pt}\selectfont 46 BUS images, 5-fold CV}} \\
\hhline{-----}
%row no:18
\multicolumn{1}{|p{0.42in}}{{\fontsize{10pt}{12.0pt}\selectfont [137]}} & 
\multicolumn{1}{|p{0.68in}}{{\fontsize{10pt}{12.0pt}\selectfont 2016}} & 
\multicolumn{1}{|p{1.36in}}{{\fontsize{10pt}{12.0pt}\selectfont SVM}} & 
\multicolumn{1}{|p{0.52in}}{{\fontsize{10pt}{12.0pt}\selectfont Breast}} & 
\multicolumn{1}{|p{2.68in}|}{{\fontsize{10pt}{12.0pt}\selectfont 50 BUS images}} \\
\hhline{-----}
%row no:19
\multicolumn{1}{|p{0.42in}}{{\fontsize{10pt}{12.0pt}\selectfont [138]}} & 
\multicolumn{1}{|p{0.68in}}{{\fontsize{10pt}{12.0pt}\selectfont 2018}} & 
\multicolumn{1}{|p{1.36in}}{{\fontsize{10pt}{12.0pt}\selectfont ConvEDNet}} & 
\multicolumn{1}{|p{0.52in}}{{\fontsize{10pt}{12.0pt}\selectfont Breast}} & 
\multicolumn{1}{|p{2.68in}|}{{\fontsize{10pt}{12.0pt}\selectfont N/A}} \\
\hhline{-----}
%row no:20
\multicolumn{1}{|p{0.42in}}{{\fontsize{10pt}{12.0pt}\selectfont [139]}} & 
\multicolumn{1}{|p{0.68in}}{{\fontsize{10pt}{12.0pt}\selectfont 2019}} & 
\multicolumn{1}{|p{1.36in}}{{\fontsize{10pt}{12.0pt}\selectfont Conditional GAN}} & 
\multicolumn{1}{|p{0.52in}}{{\fontsize{10pt}{12.0pt}\selectfont Breast}} & 
\multicolumn{1}{|p{2.68in}|}{{\fontsize{10pt}{12.0pt}\selectfont 250 images, 70/10/20}} \\
\hhline{-----}
%row no:21
\multicolumn{1}{|p{0.42in}}{{\fontsize{10pt}{12.0pt}\selectfont [140]}} & 
\multicolumn{1}{|p{0.68in}}{{\fontsize{10pt}{12.0pt}\selectfont 2019}} & 
\multicolumn{1}{|p{1.36in}}{{\fontsize{10pt}{12.0pt}\selectfont Dilated FCN}} & 
\multicolumn{1}{|p{0.52in}}{{\fontsize{10pt}{12.0pt}\selectfont Breast}} & 
\multicolumn{1}{|p{2.68in}|}{{\fontsize{10pt}{12.0pt}\selectfont 89 subjects, 570 images, 10-fold CV}} \\
\hhline{-----}
%row no:22
\multicolumn{1}{|p{0.42in}}{{\fontsize{10pt}{12.0pt}\selectfont [141]}} & 
\multicolumn{1}{|p{0.68in}}{{\fontsize{10pt}{12.0pt}\selectfont 2018}} & 
\multicolumn{1}{|p{1.36in}}{{\fontsize{10pt}{12.0pt}\selectfont FCNs}} & 
\multicolumn{1}{|p{0.52in}}{{\fontsize{10pt}{12.0pt}\selectfont Breast}} & 
\multicolumn{1}{|p{2.68in}|}{{\fontsize{10pt}{12.0pt}\selectfont 113 malignant, 356 benign lesions, 70/10/20}} \\
\hhline{-----}
%row no:23
\multicolumn{1}{|p{0.42in}}{{\fontsize{10pt}{12.0pt}\selectfont [142]}} & 
\multicolumn{1}{|p{0.68in}}{{\fontsize{10pt}{12.0pt}\selectfont 2018}} & 
\multicolumn{1}{|p{1.36in}}{{\fontsize{10pt}{12.0pt}\selectfont R-CNN}} & 
\multicolumn{1}{|p{0.52in}}{{\fontsize{10pt}{12.0pt}\selectfont Breast}} & 
\multicolumn{1}{|p{2.68in}|}{{\fontsize{10pt}{12.0pt}\selectfont 5624 images}} \\
\hhline{-----}

\end{longtable}}

%%%%%%%%%%%%%%%%%%%% Table No: 4 ends here %%%%%%%%%%%%%%%%%%%%

%\vspace{\baselineskip}
%\begin{enumerate}[label*={\fontsize{11pt}{11pt}\selectfont \arabic*.}]
	%\item \begin{enumerate}[label*={\fontsize{11pt}{11pt}\selectfont \arabic*.}]
	%\item \begin{enumerate}[label*={\fontsize{11pt}{11pt}\selectfont \arabic*.}]
	%\item \begin{enumerate}[label*={\fontsize{11pt}{11pt}\selectfont \arabic*.}]
	%\item {\fontsize{11pt}{13.2pt}\selectfont Echocardiography\par}\par
\paragraph{Echocardiography}
\begin{justify}
{\fontsize{11pt}{13.2pt}\selectfont Table 5 shows a list of selected references that used AI in echocardiography segmentation. In 2012, Carneiro \textit{et al}. developed a two-stage DL technique based on a maximum \textit{a} \textit{posteriori} framework for left ventricle (LV) segmentation: (1) selecting region of interest (ROI) in the test image in which the LV is fully present; (2) extracting the LV contour from the previous selected ROI [143]. A deep belief net (DBN) was used in this study. An average Hausdorff distance of approximately 18 mm and average mean absolute distance of 8 mm for the LV segmentation were obtained based on training dataset of 400 images from 12 patient sequences and test dataset of 50 images from 2 healthy subject sequences. In order to reduce the manual annotation, Smistad \textit{et al}. trained a U-Net model based on 2D ultrasound image frames of which the label images were obtained using an automatic Kalman filter (KF) based segmentation method [144]. Based on a manually segmented test dataset, the evaluation metrics such as DSC showed comparable performance between U-Net and KF while for the Hausdorff distance the U-Net was better than KF. This group also introduced the Cardiac Acquisitions for Multi-structure Ultrasound Segmentation (CAMUS) dataset in 2018, which was the largest publicly-available and fully-annotated dataset for assessing echocardiography [145].\  This dataset contains two and four-chamber acquisitions from 500 patients with manual contouring from cardiologists and on a fold of 50 patients. U-Net trained on this dataset produced highly accurate segmentation results of end-diastolic and end-systolic LV volumes with a mean correlation of 0.95 and an absolute mean error of 9.5 ml as well as the ejection of fraction with a mean correlation coefficient of 0.8 and absolute mean error of 5.6$\%$ . \par}
\end{justify}\par

\begin{justify}
{\fontsize{11pt}{13.2pt}\selectfont For the sake of evaluating ventricular volume and ejection fraction, LV segmentation based on three-dimensional echocardiography (3DE) is needed. Lempitsky \textit{et al}. used a random forest classifier for automatic delineation of myocardium in real-time 3DE were performed [146]. Fourteen 3D echocardiograms for left and right ventricles delineated by an expert were used for training the model (11 for training and 3 for validation). Based on the validation dataset, the parameters of the random forest classifier in terms of number of training samples for each tree N\textsubscript{train} was 100000, the maximal tree depth D\textsubscript{max} was 16, the number of random binary tests generated for each node N\textsubscript{tests }was 30, and the radius of the neighborhood R was 32. This technique achieved delineation accuracy of 92$\%$  true positive rate at 8$\%$  false positive rate. Moreover, GPU could make this technique in real-time for the full 3D volume. Suyu \textit{et al}. proposed a fully learning framework to estimate the LV volume in 3DE [147]. This framework used deep belief nets (DBN) to capture features from original volumes and trained the random forest model to estimate the LV volume. The results in terms of the correlation (R value) between the prediction and the ground truth of this framework were 0.85 for end diastole volume, 0.87 for end systole volume and 0.86 for ejection fraction. The same group in 2016 also used DL and snake model to segment LV endocardium based on 3DE [148]. Snake model is an active contour model, which looks onto edges and accurately localizes them [149]. The performance of this technique was evaluated on the Challenge dataset on Endocardial Three-dimensional Ultrasound Segmentation (CETUS) and shown i) modified dice similarity index of 0.112 in end diastole (ED) and 0.16 in end systole (ES); ii) Hausdorff distance of 8.34 mm (ED) and 8.46 mm (ES) and iii) mean surface distance of 2.2 mm (ED) and 2.56 mm (ES). Recently, Oktay \textit{et al}. developed an anatomically constrained neural network (ACNN) for 3D LV segmentation [150]. The nonlinear representation of the underlying anatomy derived from an auto-encoder network was used to constrain the network. Based on the CETUS dataset, the performance of this technique was assessed in terms of average Dice values and were 0.912 (ED) and 0.873 (ES); ii) average Hausdorff distance of 7 mm (ED) and 7.7 mm (ES) and iii) average mean absolute distances of 1.9 mm (ED) and 2.1 mm (ES), respectively. Suyu \textit{et al}.\  in 2018 proposed a real-time framework, VoxelAtlasGAN, for 3D LV segmentation on 3DE [151]. This framework utilized the voxel-to-voxel conditional GAN to fuse substantial 3D spatial context information from 3DE and embedded 3D LV atlas prior knowledge into an end-to-end optimization framework. Discrimination loss and consistent constraint were combined and used as the final optimization objective to improve the generalization of the framework. Obtained results showed the mean DSC value of 0.953, mean Hausdorff distance of 7.26 mm, mean surface distance of 1.85 mm, correlation of ejection fraction (EF) of 0.92. \par}
\end{justify}\par

%\end{enumerate}
%\end{enumerate}
%\end{enumerate}
%\end{enumerate}\subsubsection*{Table 5. Overview of AI applications in echocardiography segmentation.}
%\addcontentsline{toc}{subsubsection}{Table 5. Overview of AI applications in echocardiography segmentation.}

\begin{justify}
{\fontsize{11pt}{13.2pt}\selectfont \textbf{Table 5.} Overview of AI applications in echocardiography segmentation.\par}
\end{justify}\par
%%%%%%%%%%%%%%%%%%%% Table No: 5 starts here %%%%%%%%%%%%%%%%%%%%

\begin{table}[H]
 			\centering
\begin{tabular}{p{0.42in}p{0.55in}p{1.24in}p{3.55in}}
\hline
%row no:1
\multicolumn{1}{|p{0.42in}}{{\fontsize{11pt}{13.2pt}\selectfont \textbf{Ref}}} & 
\multicolumn{1}{|p{0.55in}}{{\fontsize{11pt}{13.2pt}\selectfont \textbf{Year}}} & 
\multicolumn{1}{|p{1.24in}}{{\fontsize{11pt}{13.2pt}\selectfont \textbf{Network}}} & 
\multicolumn{1}{|p{3.55in}|}{{\fontsize{11pt}{13.2pt}\selectfont \textbf{$\#$  of samples in training/testing datasets}}} \\
\hhline{----}
%row no:2
\multicolumn{1}{|p{0.42in}}{{\fontsize{11pt}{13.2pt}\selectfont [143]}} & 
\multicolumn{1}{|p{0.55in}}{{\fontsize{11pt}{13.2pt}\selectfont 2011}} & 
\multicolumn{1}{|p{1.24in}}{{\fontsize{11pt}{13.2pt}\selectfont DBN}} & 
\multicolumn{1}{|p{3.55in}|}{{\fontsize{11pt}{13.2pt}\selectfont 400/50 images}} \\
\hhline{----}
%row no:3
\multicolumn{1}{|p{0.42in}}{{\fontsize{11pt}{13.2pt}\selectfont [144]}} & 
\multicolumn{1}{|p{0.55in}}{{\fontsize{11pt}{13.2pt}\selectfont 2017}} & 
\multicolumn{1}{|p{1.24in}}{{\fontsize{11pt}{13.2pt}\selectfont U-Net}} & 
\multicolumn{1}{|p{3.55in}|}{{\fontsize{11pt}{13.2pt}\selectfont 87/13 subjects}} \\
\hhline{----}
%row no:4
\multicolumn{1}{|p{0.42in}}{{\fontsize{11pt}{13.2pt}\selectfont [145]}} & 
\multicolumn{1}{|p{0.55in}}{{\fontsize{11pt}{13.2pt}\selectfont 2019}} & 
\multicolumn{1}{|p{1.24in}}{{\fontsize{11pt}{13.2pt}\selectfont U-Net}} & 
\multicolumn{1}{|p{3.55in}|}{{\fontsize{11pt}{13.2pt}\selectfont 500/50 subjects}} \\
\hhline{----}
%row no:5
\multicolumn{1}{|p{0.42in}}{{\fontsize{11pt}{13.2pt}\selectfont [146]}} & 
\multicolumn{1}{|p{0.55in}}{{\fontsize{11pt}{13.2pt}\selectfont 2009}} & 
\multicolumn{1}{|p{1.24in}}{{\fontsize{11pt}{13.2pt}\selectfont Random forest }} & 
\multicolumn{1}{|p{3.55in}|}{{\fontsize{11pt}{13.2pt}\selectfont 11/3 subjects}} \\
\hhline{----}
%row no:6
\multicolumn{1}{|p{0.42in}}{{\fontsize{11pt}{13.2pt}\selectfont [147]}} & 
\multicolumn{1}{|p{0.55in}}{{\fontsize{11pt}{13.2pt}\selectfont 2016}} & 
\multicolumn{1}{|p{1.24in}}{{\fontsize{11pt}{13.2pt}\selectfont DBN}} & 
\multicolumn{1}{|p{3.55in}|}{{\fontsize{11pt}{13.2pt}\selectfont 60/60 subjects}} \\
\hhline{----}
%row no:7
\multicolumn{1}{|p{0.42in}}{{\fontsize{11pt}{13.2pt}\selectfont [148]}} & 
\multicolumn{1}{|p{0.55in}}{{\fontsize{11pt}{13.2pt}\selectfont 2016}} & 
\multicolumn{1}{|p{1.24in}}{{\fontsize{11pt}{13.2pt}\selectfont CNN}} & 
\multicolumn{1}{|p{3.55in}|}{{\fontsize{11pt}{13.2pt}\selectfont 15/30 volumes}} \\
\hhline{----}
%row no:8
\multicolumn{1}{|p{0.42in}}{{\fontsize{11pt}{13.2pt}\selectfont [150]}} & 
\multicolumn{1}{|p{0.55in}}{{\fontsize{11pt}{13.2pt}\selectfont 2017}} & 
\multicolumn{1}{|p{1.24in}}{{\fontsize{11pt}{13.2pt}\selectfont ACNN}} & 
\multicolumn{1}{|p{3.55in}|}{{\fontsize{11pt}{13.2pt}\selectfont 900/200 images}} \\
\hhline{----}
%row no:9
\multicolumn{1}{|p{0.42in}}{{\fontsize{11pt}{13.2pt}\selectfont [151]}} & 
\multicolumn{1}{|p{0.55in}}{{\fontsize{11pt}{13.2pt}\selectfont 2018}} & 
\multicolumn{1}{|p{1.24in}}{{\fontsize{11pt}{13.2pt}\selectfont VoxelAtlasGAN}} & 
\multicolumn{1}{|p{3.55in}|}{{\fontsize{11pt}{13.2pt}\selectfont 25/35 subjects}} \\
\hhline{----}

\end{tabular}
 \end{table}

%%%%%%%%%%%%%%%%%%%% Table No: 5 ends here %%%%%%%%%%%%%%%%%%%%

\vspace{\baselineskip}
\subsubsection{Registration}
{\fontsize{11pt}{13.2pt}\selectfont Multi-model registration between CT/MRI and ultrasound is challenging given the imaging characteristics of ultrasound cannot be matched with the conventional similarity metrics. Table 6 shows a list of selected references that used AI in registration. Salehi \textit{et al}. used CNN to register intra-operative ultrasound images to a reference pre-operative CT volume [152]. They developed a technique to calibrate speed of sound by focusing on the bone site and improve the imaging quality of steered compound images. Hu \textit{et al}. proposed an end-to-end CNN method for aligning multiple labelled structures for image pairs in terms of displacement field prediction [153]. This group later proposed a weakly-supervised, label-driven formulation for deriving 3D voxel correspondence from high-level label correspondence [154]. Sun \textit{et al}. used a fully CNN to train on a set of artificially generated displacement vectors for direct estimation of the displacement between a pair of multi-modal image patches [155]. However, the obtained results showed that this technique did not work on the real CT and US images, only worked on the simulated data. Prostate HDR brachytherapy could benefit significantly from MRI-targeted, ultrasound-guided procedure if MRI-defined dominant intraprostatic lesions can be incorporated into real-time ultrasound imaging to guide catheter placement. However, accurate MRI-US image registration remains a challenging task due to different gray-level intensity, image field size between MRI and US. In order to accurately track the intra-operative tissue shift, Sun \textit{et al}. used 3D CNN for non-rigid MRI-ultrasound registration [156]. This architecture is consisted of feature extractor, deformation field generator and spatial sampler. Zeng \textit{et al.} proposed a DL-based method to perform automatic MRI-US registration using weakly-supervised learning [51, 157]. They combined CNN with long short-term memory (LSTM) neural network to conduct DL-based deformation field prediction. The obtained results were a mean target registration error (TRE) of 2.85 $ \pm $  1.72 mm and a mean Dice of 0.89. \par}\par

{\fontsize{11pt}{13.2pt}\selectfont \textbf{Table 6.} Overview of AI applications in US registration. \par}\par

%%%%%%%%%%%%%%%%%%%% Table No: 6 starts here %%%%%%%%%%%%%%%%%%%%

\begin{table}[H]
 			\centering
\begin{tabular}{p{0.55in}p{0.74in}p{0.8in}p{0.52in}p{2.99in}}
\hline
%row no:1
\multicolumn{1}{|p{0.55in}}{{\fontsize{11pt}{13.2pt}\selectfont \textbf{Ref}}} & 
\multicolumn{1}{|p{0.74in}}{{\fontsize{11pt}{13.2pt}\selectfont \textbf{Year}}} & 
\multicolumn{1}{|p{0.8in}}{{\fontsize{11pt}{13.2pt}\selectfont \textbf{Network}}} & 
\multicolumn{1}{|p{0.52in}}{{\fontsize{11pt}{13.2pt}\selectfont \textbf{ROI}}} & 
\multicolumn{1}{|p{2.99in}|}{{\fontsize{11pt}{13.2pt}\selectfont \textbf{$\#$  of samples in training/testing datasets}}} \\
\hhline{-----}
%row no:2
\multicolumn{1}{|p{0.55in}}{{\fontsize{11pt}{13.2pt}\selectfont [152]}} & 
\multicolumn{1}{|p{0.74in}}{{\fontsize{11pt}{13.2pt}\selectfont 2017}} & 
\multicolumn{1}{|p{0.8in}}{{\fontsize{11pt}{13.2pt}\selectfont CNN}} & 
\multicolumn{1}{|p{0.52in}}{{\fontsize{11pt}{13.2pt}\selectfont Bone}} & 
\multicolumn{1}{|p{2.99in}|}{{\fontsize{11pt}{13.2pt}\selectfont 2-fold CV}} \\
\hhline{-----}
%row no:3
\multicolumn{1}{|p{0.55in}}{{\fontsize{11pt}{13.2pt}\selectfont [153]}} & 
\multicolumn{1}{|p{0.74in}}{{\fontsize{11pt}{13.2pt}\selectfont 2018}} & 
\multicolumn{1}{|p{0.8in}}{{\fontsize{11pt}{13.2pt}\selectfont CNN}} & 
\multicolumn{1}{|p{0.52in}}{{\fontsize{11pt}{13.2pt}\selectfont Prostate}} & 
\multicolumn{1}{|p{2.99in}|}{{\fontsize{11pt}{13.2pt}\selectfont 12-fold CV}} \\
\hhline{-----}
%row no:4
\multicolumn{1}{|p{0.55in}}{{\fontsize{11pt}{13.2pt}\selectfont [154]}} & 
\multicolumn{1}{|p{0.74in}}{{\fontsize{11pt}{13.2pt}\selectfont 2018}} & 
\multicolumn{1}{|p{0.8in}}{{\fontsize{11pt}{13.2pt}\selectfont 3D CNN}} & 
\multicolumn{1}{|p{0.52in}}{{\fontsize{11pt}{13.2pt}\selectfont Prostate}} & 
\multicolumn{1}{|p{2.99in}|}{{\fontsize{11pt}{13.2pt}\selectfont 10-fold CV}} \\
\hhline{-----}
%row no:5
\multicolumn{1}{|p{0.55in}}{{\fontsize{11pt}{13.2pt}\selectfont [155]}} & 
\multicolumn{1}{|p{0.74in}}{{\fontsize{11pt}{13.2pt}\selectfont 2018}} & 
\multicolumn{1}{|p{0.8in}}{{\fontsize{11pt}{13.2pt}\selectfont CNN}} & 
\multicolumn{1}{|p{0.52in}}{{\fontsize{11pt}{13.2pt}\selectfont Liver }} & 
\multicolumn{1}{|p{2.99in}|}{{\fontsize{11pt}{13.2pt}\selectfont N/A}} \\
\hhline{-----}
%row no:6
\multicolumn{1}{|p{0.55in}}{{\fontsize{11pt}{13.2pt}\selectfont [156]}} & 
\multicolumn{1}{|p{0.74in}}{{\fontsize{11pt}{13.2pt}\selectfont 2018}} & 
\multicolumn{1}{|p{0.8in}}{{\fontsize{11pt}{13.2pt}\selectfont 3D CNN}} & 
\multicolumn{1}{|p{0.52in}}{{\fontsize{11pt}{13.2pt}\selectfont Prostate }} & 
\multicolumn{1}{|p{2.99in}|}{{\fontsize{11pt}{13.2pt}\selectfont N/A}} \\
\hhline{-----}
%row no:7
\multicolumn{1}{|p{0.55in}}{{\fontsize{11pt}{13.2pt}\selectfont [51, 157]}} & 
\multicolumn{1}{|p{0.74in}}{{\fontsize{11pt}{13.2pt}\selectfont 2019, 2020}} & 
\multicolumn{1}{|p{0.8in}}{{\fontsize{11pt}{13.2pt}\selectfont CNN, LSTM}} & 
\multicolumn{1}{|p{0.52in}}{{\fontsize{11pt}{13.2pt}\selectfont Prostate }} & 
\multicolumn{1}{|p{2.99in}|}{{\fontsize{11pt}{13.2pt}\selectfont 36 subjects}} \\
\hhline{-----}

\end{tabular}
 \end{table}

%%%%%%%%%%%%%%%%%%%% Table No: 6 ends here %%%%%%%%%%%%%%%%%%%%

\vspace{\baselineskip}
\subsubsection{Needle detection}
\begin{justify}
{\fontsize{11pt}{13.2pt}\selectfont Minimally invasive procedure, in which a needle is inserted into internal organs, has widely been used in biopsy, brachytherapy, etc. Ultrasound imaging has broadly been used for visualizing the needle and guiding the intervention. In this procedure, accuracy of placing the needle tip into the target region is critical as failure to do this may cause serious complications and affect the treatment efficacy. However, given the low signal-to-noise ratio and artifacts in US imaging, it is a great challenge to accurately detect needle in this imaging modality. Hence, there is a great need to develop automated technique for this task. Table 7 shows a list of selected references that used AI in needle detection. Geraldes \textit{et al}. proposed to combine a multiple layer perceptron network with Kalman filter for needle detection in the ultrasound image [158]. Belgi \textit{et al}. used SVM and time-domain features for identifying invisible needle in ultrasound imaging [159]. Pourtaherian \textit{et al}. used patch classification and semantic segmentation methods for localizing needle from orthogonal 2D ultrasound images [160]. The same group\ later proposed to use  a dilated CNN for identifying partially visible needles in 3D ultrasound [161]. Mwikirize \textit{et al}. combined a fully convolutional network and a region-based CNN for detecting needle in 2D US images [162]. This group later used a digital subtraction scheme for low-level intensity enhancement and a deep learning scheme for needle tip detection [163]. In order to quantify dose distribution to adaptively guide needle displacement in real-time, a fast and accurate multi-needle reconstruction algorithm is needed. However, due to low SNR and image artifacts inherited in US imaging, automatic multiple catheter reconstruction in US images is still a challenging. Zhang \textit{et al}. proposed an unsupervised dictionary learning technique to quickly and accurately reconstruct multiple catheters in 3D US images with needles as labels while images without needles as auxiliary [164].\  A sparse dictionary learning technique coupled with order-graph regularized dictionary learning was implemented. This group later used a U-net architecture incorporate attention gates to segment multiple needles in the 3D TRUS images of HDR prostate brachytherapy [165]. This technique was able to detect 96$\%$  needles with 0.29 $ \pm $  0.236 mm at shaft error and 0.442 $ \pm $  0.831 mm at tip error. This group also proposed a workflow for multiple needle detection in 3D US images with corresponding CT images as ground truth [166]. Given the CT and US images do not match exactly, they proposed a dubbed Bidirectional Convolutional Sparse Coding (BiCSC) scheme which extract the latent features from US and CT and then map the learned features from US to the features from CT, to tackle this weakly supervised problem. A clustering algorithm was used to locate the needle position and the random sample consensus algorithm was used to model the needle in the ROI. \par}
\end{justify}\par

\vspace{\baselineskip}
{\fontsize{11pt}{13.2pt}\selectfont \textbf{Table 7.} Overview of AI applications in US needle detection.\par}\par

%%%%%%%%%%%%%%%%%%%% Table No: 7 starts here %%%%%%%%%%%%%%%%%%%%

\begin{table}[H]
 			\centering
\begin{tabular}{p{0.36in}p{0.31in}p{1.3in}p{1.05in}p{2.59in}}
\hline
%row no:1
\multicolumn{1}{|p{0.36in}}{{\fontsize{11pt}{13.2pt}\selectfont \textbf{Ref}}} & 
\multicolumn{1}{|p{0.31in}}{{\fontsize{11pt}{13.2pt}\selectfont \textbf{Year}}} & 
\multicolumn{1}{|p{1.3in}}{{\fontsize{11pt}{13.2pt}\selectfont \textbf{Network}}} & 
\multicolumn{1}{|p{1.05in}}{{\fontsize{11pt}{13.2pt}\selectfont \textbf{ROI}}} & 
\multicolumn{1}{|p{2.59in}|}{{\fontsize{11pt}{13.2pt}\selectfont \textbf{$\#$  of samples in training/testing datasets}}} \\
\hhline{-----}
%row no:2
\multicolumn{1}{|p{0.36in}}{{\fontsize{11pt}{13.2pt}\selectfont [158]}} & 
\multicolumn{1}{|p{0.31in}}{{\fontsize{11pt}{13.2pt}\selectfont 2014}} & 
\multicolumn{1}{|p{1.3in}}{{\fontsize{11pt}{13.2pt}\selectfont MLP}} & 
\multicolumn{1}{|p{1.05in}}{{\fontsize{11pt}{13.2pt}\selectfont Phantom}} & 
\multicolumn{1}{|p{2.59in}|}{{\fontsize{11pt}{13.2pt}\selectfont 1335/422 frames}} \\
\hhline{-----}
%row no:3
\multicolumn{1}{|p{0.36in}}{{\fontsize{11pt}{13.2pt}\selectfont [159]}} & 
\multicolumn{1}{|p{0.31in}}{{\fontsize{11pt}{13.2pt}\selectfont 2017}} & 
\multicolumn{1}{|p{1.3in}}{{\fontsize{11pt}{13.2pt}\selectfont SVM}} & 
\multicolumn{1}{|p{1.05in}}{{\fontsize{11pt}{13.2pt}\selectfont Porcine muscle }} & 
\multicolumn{1}{|p{2.59in}|}{{\fontsize{11pt}{13.2pt}\selectfont 20 sequences, 50/25/25 }} \\
\hhline{-----}
%row no:4
\multicolumn{1}{|p{0.36in}}{{\fontsize{11pt}{13.2pt}\selectfont [160]}} & 
\multicolumn{1}{|p{0.31in}}{{\fontsize{11pt}{13.2pt}\selectfont 2018}} & 
\multicolumn{1}{|p{1.3in}}{} & 
\multicolumn{1}{|p{1.05in}}{{\fontsize{11pt}{13.2pt}\selectfont Chicken breast, porcine leg}} & 
\multicolumn{1}{|p{2.59in}|}{{\fontsize{11pt}{13.2pt}\selectfont 4/1 subset}} \\
\hhline{-----}
%row no:5
\multicolumn{1}{|p{0.36in}}{{\fontsize{11pt}{13.2pt}\selectfont [161]}} & 
\multicolumn{1}{|p{0.31in}}{{\fontsize{11pt}{13.2pt}\selectfont 2018}} & 
\multicolumn{1}{|p{1.3in}}{{\fontsize{11pt}{13.2pt}\selectfont Dilated CNN}} & 
\multicolumn{1}{|p{1.05in}}{{\fontsize{11pt}{13.2pt}\selectfont 3D volumes}} & 
\multicolumn{1}{|p{2.59in}|}{{\fontsize{11pt}{13.2pt}\selectfont 20 datasets}} \\
\hhline{-----}
%row no:6
\multicolumn{1}{|p{0.36in}}{{\fontsize{11pt}{13.2pt}\selectfont [162]}} & 
\multicolumn{1}{|p{0.31in}}{{\fontsize{11pt}{13.2pt}\selectfont 2018}} & 
\multicolumn{1}{|p{1.3in}}{{\fontsize{11pt}{13.2pt}\selectfont FCN, R-CNN}} & 
\multicolumn{1}{|p{1.05in}}{{\fontsize{11pt}{13.2pt}\selectfont Bovine, porcine ex vivo}} & 
\multicolumn{1}{|p{2.59in}|}{{\fontsize{11pt}{13.2pt}\selectfont 2500 scans}} \\
\hhline{-----}
%row no:7
\multicolumn{1}{|p{0.36in}}{{\fontsize{11pt}{13.2pt}\selectfont [163]}} & 
\multicolumn{1}{|p{0.31in}}{{\fontsize{11pt}{13.2pt}\selectfont 2019}} & 
\multicolumn{1}{|p{1.3in}}{{\fontsize{11pt}{13.2pt}\selectfont DNN}} & 
\multicolumn{1}{|p{1.05in}}{{\fontsize{11pt}{13.2pt}\selectfont Phantoms }} & 
\multicolumn{1}{|p{2.59in}|}{{\fontsize{11pt}{13.2pt}\selectfont 5000/1000 images}} \\
\hhline{-----}
%row no:8
\multicolumn{1}{|p{0.36in}}{{\fontsize{11pt}{13.2pt}\selectfont [164]}} & 
\multicolumn{1}{|p{0.31in}}{{\fontsize{11pt}{13.2pt}\selectfont 2020}} & 
\multicolumn{1}{|p{1.3in}}{{\fontsize{11pt}{13.2pt}\selectfont Unsupervised dictionary learning}} & 
\multicolumn{1}{|p{1.05in}}{{\fontsize{11pt}{13.2pt}\selectfont Prostate }} & 
\multicolumn{1}{|p{2.59in}|}{{\fontsize{11pt}{13.2pt}\selectfont 70/21 patients}} \\
\hhline{-----}
%row no:9
\multicolumn{1}{|p{0.36in}}{{\fontsize{11pt}{13.2pt}\selectfont [165]}} & 
\multicolumn{1}{|p{0.31in}}{{\fontsize{11pt}{13.2pt}\selectfont 2020}} & 
\multicolumn{1}{|p{1.3in}}{{\fontsize{11pt}{13.2pt}\selectfont U-net with attention gates}} & 
\multicolumn{1}{|p{1.05in}}{{\fontsize{11pt}{13.2pt}\selectfont Prostate }} & 
\multicolumn{1}{|p{2.59in}|}{{\fontsize{11pt}{13.2pt}\selectfont 23 patients}} \\
\hhline{-----}
%row no:10
\multicolumn{1}{|p{0.36in}}{{\fontsize{11pt}{13.2pt}\selectfont [166]}} & 
\multicolumn{1}{|p{0.31in}}{{\fontsize{11pt}{13.2pt}\selectfont 2020}} & 
\multicolumn{1}{|p{1.3in}}{{\fontsize{11pt}{13.2pt}\selectfont CNN}} & 
\multicolumn{1}{|p{1.05in}}{{\fontsize{11pt}{13.2pt}\selectfont Prostate }} & 
\multicolumn{1}{|p{2.59in}|}{{\fontsize{11pt}{13.2pt}\selectfont 10 patients}} \\
\hhline{-----}

\end{tabular}
 \end{table}

%%%%%%%%%%%%%%%%%%%% Table No: 7 ends here %%%%%%%%%%%%%%%%%%%%

%\vspace{\baselineskip}
%\begin{enumerate}[label*={\fontsize{11pt}{11pt}\selectfont \textbf{\arabic*.}}]
	%\item \begin{enumerate}[label*={\fontsize{11pt}{11pt}\selectfont \textbf{\arabic*.}}]
	%\item \begin{enumerate}[label*={\fontsize{11pt}{11pt}\selectfont \textbf{\arabic*.}}]
	%\item {\fontsize{11pt}{13.2pt}\selectfont \textbf{Image synthesis}\par}\par
\subsubsection{Image synthesis}
\begin{justify}
{\fontsize{11pt}{13.2pt}\selectfont Table 8 shows a list of selected references that used AI in image synthesis. Tom \textit{et al}. proposed a stacked GAN for simulating patho-realistic intravascular ultrasound images (IVUS) using a two stage GAN architecture [167]. The stage I GAN preserve tissue specific speckle intensities while stage II GAN generated high resolution images with patho-realistic speckle profiles. Visual Turing test and shift in tissue specific intensity distributions of the images were used to evaluate the degree of similarity between real and simulated images. Hu \textit{et al}. suggested to use conditional GANs to simulate ultrasound images at given 3D spatial locations acquired from freehand ultrasound transducer [168]. Calibrated pixel coordinates in global physical space were taken as conditional input into the network. Residual network units and shortcuts of conditioning data were used in the discriminator and generator of conditional GAN. Fujioka \textit{et al}. used deep convolutional GAN (DCGANs) to generate breast ultrasound images [169]. Images from both the benign and malignant masses in the breasts were included as input. The DCGAN discriminator consists of stride convolution layers, batch norm layers and LeakyReLU activations while generator is composed of transpose-convolutional layers, batch norm layers and LeakyReLU activations. The\ results showed that the generated high-quality breast ultrasound images were indistinguishable from the original images from clinical perspective.  \par}
\end{justify}\par

\begin{justify}
{\fontsize{11pt}{13.2pt}\selectfont \textbf{Table 8.} Overview of AI applications in US image synthesis.\par}
\end{justify}\par

%%%%%%%%%%%%%%%%%%%% Table No: 8 starts here %%%%%%%%%%%%%%%%%%%%

\begin{table}[H]
 			\centering
\begin{tabular}{p{0.3in}p{0.3in}p{1.11in}p{0.55in}p{3.18in}}
\hline
%row no:1
\multicolumn{1}{|p{0.3in}}{{\fontsize{11pt}{13.2pt}\selectfont \textbf{Ref}}} & 
\multicolumn{1}{|p{0.3in}}{{\fontsize{11pt}{13.2pt}\selectfont \textbf{Year}}} & 
\multicolumn{1}{|p{1.11in}}{{\fontsize{11pt}{13.2pt}\selectfont \textbf{Network}}} & 
\multicolumn{1}{|p{0.55in}}{{\fontsize{11pt}{13.2pt}\selectfont \textbf{ROI}}} & 
\multicolumn{1}{|p{3.18in}|}{{\fontsize{11pt}{13.2pt}\selectfont \textbf{$\#$  of samples in training/testing datasets}}} \\
\hhline{-----}
%row no:2
\multicolumn{1}{|p{0.3in}}{{\fontsize{11pt}{13.2pt}\selectfont [167]}} & 
\multicolumn{1}{|p{0.3in}}{{\fontsize{11pt}{13.2pt}\selectfont 2018}} & 
\multicolumn{1}{|p{1.11in}}{{\fontsize{11pt}{13.2pt}\selectfont GAN}} & 
\multicolumn{1}{|p{0.55in}}{{\fontsize{11pt}{13.2pt}\selectfont Vascular}} & 
\multicolumn{1}{|p{3.18in}|}{{\fontsize{11pt}{13.2pt}\selectfont 2025/150 subjects}} \\
\hhline{-----}
%row no:3
\multicolumn{1}{|p{0.3in}}{{\fontsize{11pt}{13.2pt}\selectfont [168]}} & 
\multicolumn{1}{|p{0.3in}}{{\fontsize{11pt}{13.2pt}\selectfont 2017}} & 
\multicolumn{1}{|p{1.11in}}{{\fontsize{11pt}{13.2pt}\selectfont Conditional GAN}} & 
\multicolumn{1}{|p{0.55in}}{{\fontsize{11pt}{13.2pt}\selectfont Fetus}} & 
\multicolumn{1}{|p{3.18in}|}{{\fontsize{11pt}{13.2pt}\selectfont 26396 subjects}} \\
\hhline{-----}
%row no:4
\multicolumn{1}{|p{0.3in}}{{\fontsize{11pt}{13.2pt}\selectfont [169]}} & 
\multicolumn{1}{|p{0.3in}}{{\fontsize{11pt}{13.2pt}\selectfont 2019}} & 
\multicolumn{1}{|p{1.11in}}{{\fontsize{11pt}{13.2pt}\selectfont DCGAN}} & 
\multicolumn{1}{|p{0.55in}}{{\fontsize{11pt}{13.2pt}\selectfont Breast}} & 
\multicolumn{1}{|p{3.18in}|}{{\fontsize{11pt}{13.2pt}\selectfont 1057 subjects}} \\
\hhline{-----}

\end{tabular}
 \end{table}

%%%%%%%%%%%%%%%%%%%% Table No: 8 ends here %%%%%%%%%%%%%%%%%%%%

%\vspace{\baselineskip}

%\end{enumerate}
%\end{enumerate}
	%\item {\fontsize{11pt}{13.2pt}\selectfont \textbf{Statistics}\par}\par
\section{Statistics}
{\fontsize{11pt}{13.2pt}\selectfont We have analyzed the percentage distribution of some attributes including region of interest (ROI), architecture, dimension, ML/DL. The percentage distributions were shown in Figure 2. In terms of the ROI, prostate and breast sites are the most studied sites. The reason for the wide adoption of prostate and breast may be the advantage of AI for the cancer planning and treatment in the US imaging. The category of supervised CNN methods accounts for 62$\%$  while the category of GAN accounts for 6$\%$  of all methods. 84$\%$  of the works were solving 2D problems. Almost all 2D QUS problems used whole image-based training given it requires less memory for 2D US images than 3D US images. The 3D QUS problems were mostly related to 3D freehand US imaging and 3D echocardiography. With the supervisor performance of DL-based technique in the field of medical imaging, the number of DL-based QUS papers is increasing and accounts for 76$\%$  of the works. \par}\par

%%%%%%%%%%%%%%%%%%%% Figure/Image No: 2 starts here %%%%%%%%%%%%%%%%%%%%

\begin{figure}[H]
	\begin{Center}
		\includegraphics[width=6.5in,height=3.66in]{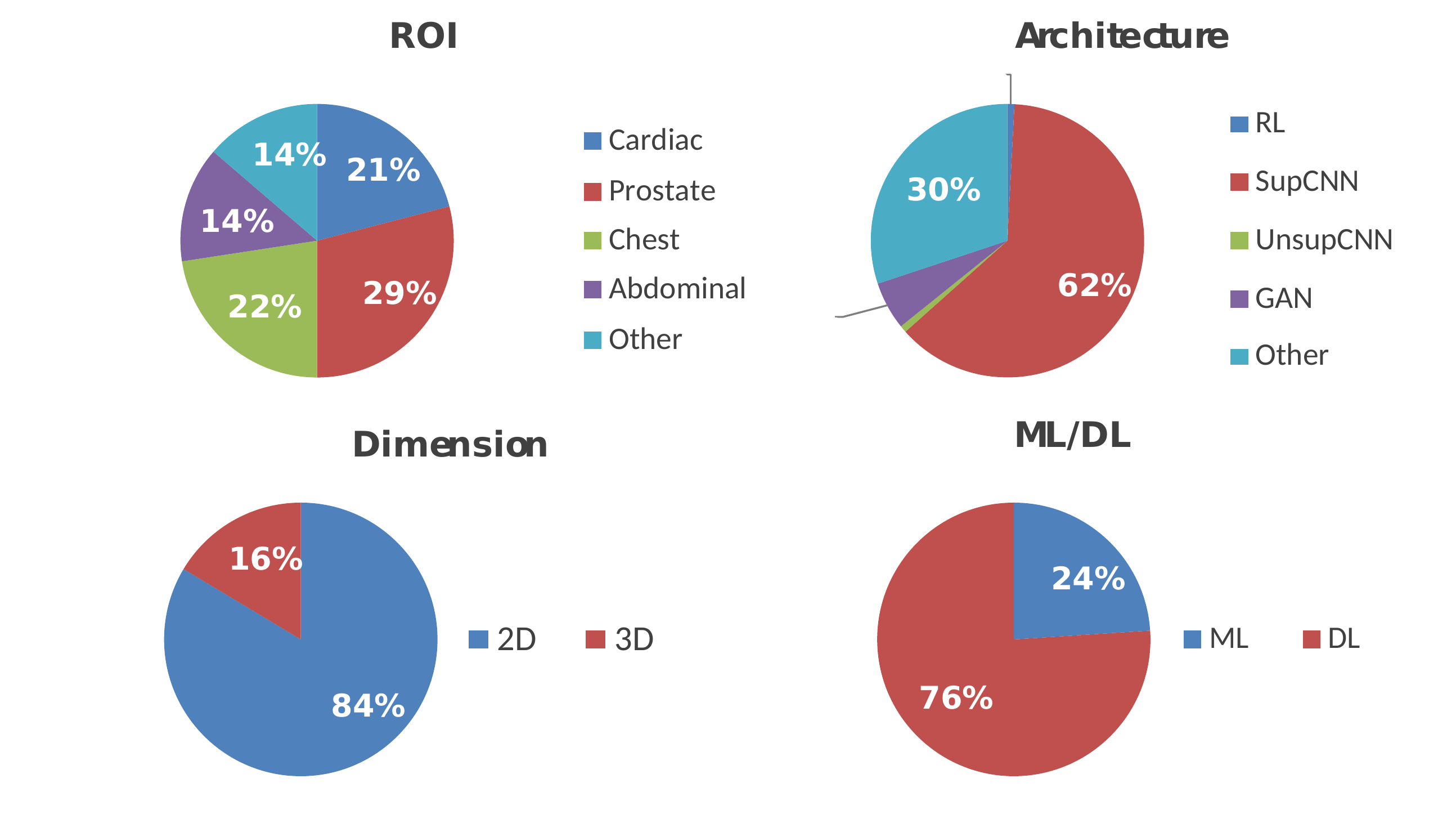}
	\end{Center}
\end{figure}

%%%%%%%%%%%%%%%%%%%% Figure/Image No: 2 Ends here %%%%%%%%%%%%%%%%%%%%

%{\fontsize{11pt}{13.2pt}\selectfont  \par}\par

%{\fontsize{10pt}{12.0pt}\selectfont \textbf{Figure 2.} Percentage pie chart of various attributes of AI-based QUS methods.\par}\par

\begin{Center}
{\fontsize{11pt}{13.2pt}\selectfont \textbf{Figure 2.} Percentage pie chart of various attributes of AI-based QUS methods.\par}
\end{Center}\par
%\end{enumerate}

\section{Challenges and potential AI applications in QUS }

\vspace{\baselineskip}
\begin{justify}
{\fontsize{11pt}{13.2pt}\selectfont Although there are numerous AI applications in QUS, certain challenges still exist. Some challenges are resulted from the poor image quality such as poor spatial resolution and high noise which are inherent in the modality of ultrasound imaging or lack of standardization of imaging acquisition parameters, such as image resolution, timing of contrast agents use, models that vary among different institutions, etc. This issue may be solved using robust AI models with data augmentation to increase the model robustness. On the other hand, medical data is not easily accessible and suffers from class imbalances. Annotation of medical images requires extensive effort and time of medical professionals, making it significantly expensive. Moreover, severe inter-operator and inter-observer variability among sonographers and physicians, which depends on the institutions’ acquisition protocols and observer preference, requires a larger training dataset to alleviate the variation. Transfer learning and weakly supervised learning have been used to address this challenge. However, several issues still exist, such as negative transfer, heterogeneous feature spaces, generalization across different tasks. For this reason, develop domain-specific AI\ models for QUS can improve the performance compared to the transfer learning pre-trained in another domain (e.g., natural images).  \par}
\end{justify}\par

\begin{justify}
{\fontsize{11pt}{13.2pt}\selectfont 3D QUS is of great importance in the field of medical imaging and has shown great potential in US-based clinical utilization. Breast tumor visualization using 3D QUS based on automated breast ultrasound technique was developed [170]. The whole patient’s breast was automatically scanned and the tumor was identified from acquired frames. AI models can be trained and used for classifying breast tumors as benign or malignant based on the parametric maps over the whole tumor volume. In another study, a 3D quantitative ultrasound elastography technique was developed for detecting prostate tumor [171]. 3D volumetric data including absolute value of tissue elasticity, strain and frequency-response were obtained. Moreover, spatiotemporal data based on video clips from QUS could be used for training 3D CNN or RNN models for real-time segmentation and quantification in QUS. \par}
\end{justify}\par

\begin{justify}
{\fontsize{11pt}{13.2pt}\selectfont \textbf{Acknowledgements}\par}
\end{justify}\par

\begin{justify}
{\fontsize{11pt}{13.2pt}\selectfont This research is supported in part by the National Cancer Institute of the National Institutes of Health under Award Number R01CA215718, the Department of Defense (DoD) Prostate Cancer Research Program (PCRP) Award W81XWH-17-1-0438 and Dunwoody Golf Club Prostate Cancer Research Award, a philanthropic award provided by the Winship Cancer Institute of Emory University.\par}
\end{justify}\par

\vspace{\baselineskip}
\begin{justify}
{\fontsize{11pt}{13.2pt}\selectfont \textbf{Conflict of Interest}\par}
\end{justify}\par

\begin{justify}
{\fontsize{11pt}{13.2pt}\selectfont The authors declare no conflicts of interest. \par}
\end{justify}\par

 %%%%%%%%%%%%  Starting New Page here %%%%%%%%%%%%%%

\newpage

\vspace{\baselineskip}\begin{justify}
{\fontsize{11pt}{13.2pt}\selectfont \textbf{References}\par}
\end{justify}\par

\setlength{\parskip}{0.0pt}
{\fontsize{11pt}{13.2pt}\selectfont 1.\tab Oelze, M.L. and J. Mamou, \textit{Review of quantitative ultrasound: Envelope statistics and backscatter coefficient imaging and contributions to diagnostic ultrasound. IEEE transactions on ultrasonics, ferroelectrics, and frequency control, 2016. \textit{63(2): p. 336-351.}}\par}\par

{\fontsize{11pt}{13.2pt}\selectfont 2.\tab Feleppa, E.J., et al. \textit{Quantitative ultrasound in cancer imaging. in Seminars in oncology. 2011. Elsevier.}\par}\par

{\fontsize{11pt}{13.2pt}\selectfont 3.\tab Litjens, G., et al., \textit{A survey on deep learning in medical image analysis. Medical image analysis, 2017. \textit{42: p. 60-88.}}\par}\par

{\fontsize{11pt}{13.2pt}\selectfont 4.\tab Kotsiantis, S.B., I.D. Zaharakis, and P.E. Pintelas, \textit{Machine learning: a review of classification and combining techniques. Artificial Intelligence Review, 2006. \textit{26(3): p. 159-190.}}\par}\par

{\fontsize{11pt}{13.2pt}\selectfont 5.\tab Shen, D., G. Wu, and H.-I. Suk, \textit{Deep learning in medical image analysis. Annual review of biomedical engineering, 2017. \textit{19: p. 221-248.}}\par}\par

{\fontsize{11pt}{13.2pt}\selectfont 6.\tab Brattain, L.J., et al., \textit{Machine learning for medical ultrasound: status, methods, and future opportunities. Abdominal Radiology, 2018. \textit{43(4): p. 786-799.}}\par}\par

{\fontsize{11pt}{13.2pt}\selectfont 7.\tab Huang, Q., F. Zhang, and X. Li, \textit{Machine learning in ultrasound computer-aided diagnostic systems: a survey. BioMed research international, 2018. \textit{2018.}}\par}\par

{\fontsize{11pt}{13.2pt}\selectfont 8.\tab van Sloun, R.J., R. Cohen, and Y.C. Eldar, \textit{Deep learning in ultrasound imaging. Proceedings of the IEEE, 2019. \textit{108(1): p. 11-29.}}\par}\par

{\fontsize{11pt}{13.2pt}\selectfont 9.\tab Zhang, H., \textit{The optimality of naive Bayes. AA, 2004. \textit{1(2): p. 3.}}\par}\par

{\fontsize{11pt}{13.2pt}\selectfont 10.\tab Suykens, J.A. and J. Vandewalle, \textit{Least squares support vector machine classifiers. Neural processing letters, 1999. \textit{9(3): p. 293-300.}}\par}\par

{\fontsize{11pt}{13.2pt}\selectfont 11.\tab Liaw, A. and M. Wiener, \textit{Classification and regression by randomForest. R news, 2002. \textit{2(3): p. 18-22.}}\par}\par

{\fontsize{11pt}{13.2pt}\selectfont 12.\tab Wang, J. and L. Perez, \textit{The effectiveness of data augmentation in image classification using deep learning. Convolutional Neural Networks Vis. Recognit, 2017: p. 11.}\par}\par

{\fontsize{11pt}{13.2pt}\selectfont 13.\tab Buslaev, A., et al., \textit{Albumentations: fast and flexible image augmentations. arXiv preprint arXiv:1809.06839, 2018.}\par}\par

{\fontsize{11pt}{13.2pt}\selectfont 14.\tab Jung, A., \textit{Imgaug: a library for image augmentation in machine learning experiments. 2017.}\par}\par

{\fontsize{11pt}{13.2pt}\selectfont 15.\tab Bloice, M.D., C. Stocker, and A. Holzinger, \textit{Augmentor: an image augmentation library for machine learning. arXiv preprint arXiv:1708.04680, 2017.}\par}\par

{\fontsize{11pt}{13.2pt}\selectfont 16.\tab Documentation, K., \textit{Image Preprocessing. Internet]. Available: \href{https://keras}{https://keras}. io/prepro cessing /image/$\#$  imagedatag enerator.}\par}\par

{\fontsize{11pt}{13.2pt}\selectfont 17.\tab Marcel, S. and Y. Rodriguez. \textit{Torchvision the machine-vision package of torch. in Proceedings of the 18th ACM international conference on Multimedia. 2010.}\par}\par

{\fontsize{11pt}{13.2pt}\selectfont 18.\tab Goodfellow, I., et al. \textit{Generative adversarial nets. in Advances in neural information processing systems. 2014.}\par}\par

{\fontsize{11pt}{13.2pt}\selectfont 19.\tab Yi, X., E. Walia, and P. Babyn, \textit{Generative adversarial network in medical imaging: A review. Medical image analysis, 2019: p. 101552.}\par}\par

{\fontsize{11pt}{13.2pt}\selectfont 20.\tab Kira, K. and L.A. Rendell, \textit{A practical approach to feature selection, in Machine Learning Proceedings 1992. 1992, Elsevier. p. 249-256.}\par}\par

{\fontsize{11pt}{13.2pt}\selectfont 21.\tab Liu, H., et al. \textit{Feature selection: An ever evolving frontier in data mining. in Feature selection in data mining. 2010.}\par}\par

{\fontsize{11pt}{13.2pt}\selectfont 22.\tab Krizhevsky, A., I. Sutskever, and G.E. Hinton. \textit{Imagenet classification with deep convolutional neural networks. in Advances in neural information processing systems. 2012.}\par}\par

{\fontsize{11pt}{13.2pt}\selectfont 23.\tab Ronneberger, O., P. Fischer, and T. Brox. \textit{U-net: Convolutional networks for biomedical image segmentation. in International Conference on Medical image computing and computer-assisted intervention. 2015. Springer.}\par}\par

{\fontsize{11pt}{13.2pt}\selectfont 24.\tab He, K., et al. \textit{Deep residual learning for image recognition. in Proceedings of the IEEE conference on computer vision and pattern recognition. 2016.}\par}\par

{\fontsize{11pt}{13.2pt}\selectfont 25.\tab Huang, G., et al. \textit{Densely connected convolutional networks. in Proceedings of the IEEE conference on computer vision and pattern recognition. 2017.}\par}\par

{\fontsize{11pt}{13.2pt}\selectfont 26.\tab Mikolov, T., et al. \textit{Recurrent neural network based language model. in Eleventh annual conference of the international speech communication association. 2010.}\par}\par

{\fontsize{11pt}{13.2pt}\selectfont 27.\tab Gregor, K., et al., \textit{Draw: A recurrent neural network for image generation. arXiv preprint arXiv:1502.04623, 2015.}\par}\par

{\fontsize{11pt}{13.2pt}\selectfont 28.\tab Sutton, R.S. and A.G. Barto, \textit{Reinforcement learning: An introduction. 2018: MIT press.}\par}\par

{\fontsize{11pt}{13.2pt}\selectfont 29.\tab Hinton, G.E. and R.R. Salakhutdinov, \textit{Reducing the dimensionality of data with neural networks. science, 2006. \textit{313(5786): p. 504-507.}}\par}\par

{\fontsize{11pt}{13.2pt}\selectfont 30.\tab Hinton, G.E., \textit{A practical guide to training restricted Boltzmann machines, in Neural networks: Tricks of the trade. 2012, Springer. p. 599-619.}\par}\par

{\fontsize{11pt}{13.2pt}\selectfont 31.\tab Deng, L. and D. Yu, \textit{Deep Learning: Methods and Applications (Foundations and Trends in Signal Processing Series Book 20). 2014, Now Publishers.}\par}\par

{\fontsize{11pt}{13.2pt}\selectfont 32.\tab Bengio, Y., A. Courville, and P. Vincent, \textit{Representation learning: A review and new perspectives. IEEE transactions on pattern analysis and machine intelligence, 2013. \textit{35(8): p. 1798-1828.}}\par}\par

{\fontsize{11pt}{13.2pt}\selectfont 33.\tab Vincent, P., et al., \textit{Stacked denoising autoencoders: Learning useful representations in a deep network with a local denoising criterion. Journal of machine learning research, 2010. \textit{11(Dec): p. 3371-3408.}}\par}\par

{\fontsize{11pt}{13.2pt}\selectfont 34.\tab Pan, S.J. and Q. Yang, \textit{A survey on transfer learning. IEEE Transactions on knowledge and data engineering, 2009. \textit{22(10): p. 1345-1359.}}\par}\par

{\fontsize{11pt}{13.2pt}\selectfont 35.\tab Dai, W., et al. \textit{Boosting for transfer learning. in Proceedings of the 24th international conference on Machine learning. 2007.}\par}\par

{\fontsize{11pt}{13.2pt}\selectfont 36.\tab Bengio, Y. \textit{Deep learning of representations for unsupervised and transfer learning. in Proceedings of ICML workshop on unsupervised and transfer learning. 2012.}\par}\par

{\fontsize{11pt}{13.2pt}\selectfont 37.\tab Zheng, Q., G. Tastan, and Y. Fan. \textit{Transfer learning for diagnosis of congenital abnormalities of the kidney and urinary tract in children based on Ultrasound imaging data. in 2018 IEEE 15th International Symposium on Biomedical Imaging (ISBI 2018). 2018. IEEE.}\par}\par

{\fontsize{11pt}{13.2pt}\selectfont 38.\tab Byra, M., et al., \textit{Transfer learning with deep convolutional neural network for liver steatosis assessment in ultrasound images. International journal of computer assisted radiology and surgery, 2018. \textit{13(12): p. 1895-1903.}}\par}\par

{\fontsize{11pt}{13.2pt}\selectfont 39.\tab Vedula, S., et al., \textit{Learning beamforming in ultrasound imaging. arXiv preprint arXiv:1812.08043, 2018.}\par}\par

{\fontsize{11pt}{13.2pt}\selectfont 40.\tab Senouf, O., et al. \textit{High frame-rate cardiac ultrasound imaging with deep learning. in International Conference on Medical Image Computing and Computer-Assisted Intervention. 2018. Springer.}\par}\par

{\fontsize{11pt}{13.2pt}\selectfont 41.\tab Perdios, D., et al. \textit{A deep learning approach to ultrasound image recovery. in 2017 IEEE International Ultrasonics Symposium (IUS). 2017. Ieee.}\par}\par

{\fontsize{11pt}{13.2pt}\selectfont 42.\tab Simson, W., et al., \textit{End-to-end learning-based ultrasound reconstruction. arXiv preprint arXiv:1904.04696, 2019.}\par}\par

{\fontsize{11pt}{13.2pt}\selectfont 43.\tab Khan, S., J. Huh, and J.C. Ye, \textit{Universal deep beamformer for variable rate ultrasound imaging. arXiv preprint arXiv:1901.01706, 2019.}\par}\par

{\fontsize{11pt}{13.2pt}\selectfont 44.\tab Nair, A.A., et al. \textit{A generative adversarial neural network for beamforming ultrasound images: Invited presentation. in 2019 53rd Annual Conference on Information Sciences and Systems (CISS). 2019. IEEE.}\par}\par

{\fontsize{11pt}{13.2pt}\selectfont 45.\tab Goudarzi, S., A. Asif, and H. Rivaz. \textit{Multi-focus ultrasound imaging using generative adversarial networks. in 2019 IEEE 16th International Symposium on Biomedical Imaging (ISBI 2019). 2019. IEEE.}\par}\par

{\fontsize{11pt}{13.2pt}\selectfont 46.\tab Yoon, Y.H., et al., \textit{Efficient b-mode ultrasound image reconstruction from sub-sampled rf data using deep learning. IEEE transactions on medical imaging, 2018. \textit{38(2): p. 325-336.}}\par}\par

{\fontsize{11pt}{13.2pt}\selectfont 47.\tab Luchies, A.C. and B.C. Byram, \textit{Deep neural networks for ultrasound beamforming. IEEE transactions on medical imaging, 2018. \textit{37(9): p. 2010-2021.}}\par}\par

{\fontsize{11pt}{13.2pt}\selectfont 48.\tab Wang, K., et al., \textit{Deep learning Radiomics of shear wave elastography significantly improved diagnostic performance for assessing liver fibrosis in chronic hepatitis B: a prospective multicentre study. Gut, 2019. \textit{68(4): p. 729-741.}}\par}\par

{\fontsize{11pt}{13.2pt}\selectfont 49.\tab Chang, R.-F., et al., \textit{Solid breast masses: neural network analysis of vascular features at three-dimensional power Doppler US for benign or malignant classification. Radiology, 2007. \textit{243(1): p. 56-62.}}\par}\par

{\fontsize{11pt}{13.2pt}\selectfont 50.\tab Ma, J., et al., \textit{A pre-trained convolutional neural network based method for thyroid nodule diagnosis. Ultrasonics, 2017. \textit{73: p. 221-230.}}\par}\par

{\fontsize{11pt}{13.2pt}\selectfont 51.\tab Zeng, Q., et al. \textit{Surface-Driven MRI-US Registration Using Weakly-Supervised Learning in Prostate Brachytherapy. in MEDICAL PHYSICS. 2019. WILEY 111 RIVER ST, HOBOKEN 07030-5774, NJ USA.}\par}\par

{\fontsize{11pt}{13.2pt}\selectfont 52.\tab Zhang, X., et al., \textit{Lung ultrasound surface wave elastography: a pilot clinical study. IEEE transactions on ultrasonics, ferroelectrics, and frequency control, 2017. \textit{64(9): p. 1298-1304.}}\par}\par

{\fontsize{11pt}{13.2pt}\selectfont 53.\tab Clay, R., et al., \textit{Assessment of Interstitial Lung Disease Using Lung Ultrasound Surface Wave Elastography. Journal of thoracic imaging, 2019. \textit{34(5): p. 313-319.}}\par}\par

{\fontsize{11pt}{13.2pt}\selectfont 54.\tab Zhang, X., et al., \textit{An ultrasound surface wave technique for assessing skin and lung diseases. Ultrasound in medicine $\&$  biology, 2018. \textit{44(2): p. 321-331.}}\par}\par

{\fontsize{11pt}{13.2pt}\selectfont 55.\tab Zhou, B., et al., \textit{Lung US surface wave elastography in interstitial lung disease staging. Radiology, 2019. \textit{291(2): p. 479-484.}}\par}\par

{\fontsize{11pt}{13.2pt}\selectfont 56.\tab Wu, S., et al. \textit{Direct reconstruction of ultrasound elastography using an end-to-end deep neural network. in International Conference on Medical Image Computing and Computer-Assisted Intervention. 2018. Springer.}\par}\par

{\fontsize{11pt}{13.2pt}\selectfont 57.\tab Gao, Z., et al., \textit{Learning the implicit strain reconstruction in ultrasound elastography using privileged information. Medical image analysis, 2019. \textit{58: p. 101534.}}\par}\par

{\fontsize{11pt}{13.2pt}\selectfont 58.\tab Ahmed, T. and M. Hasan, \textit{SHEAR-net: An End-to-End Deep Learning Approach for Single Push Ultrasound Shear Wave Elasticity Imaging. arXiv preprint arXiv:1902.04845, 2019.}\par}\par

{\fontsize{11pt}{13.2pt}\selectfont 59.\tab Feigin, M., D. Freedman, and B.W. Anthony, \textit{A deep learning framework for single-sided sound speed inversion in medical ultrasound. arXiv preprint arXiv:1810.00322, 2018.}\par}\par

{\fontsize{11pt}{13.2pt}\selectfont 60.\tab Mozaffari, M.H. and W.-S. Lee, \textit{Freehand 3-D ultrasound imaging: a systematic review. Ultrasound in medicine $\&$  biology, 2017. \textit{43(10): p. 2099-2124.}}\par}\par

{\fontsize{11pt}{13.2pt}\selectfont 61.\tab Prevost, R., et al. \textit{Deep learning for sensorless 3D freehand ultrasound imaging. in International conference on medical image computing and computer-assisted intervention. 2017. Springer.}\par}\par

{\fontsize{11pt}{13.2pt}\selectfont 62.\tab Prevost, R., et al., \textit{3D freehand ultrasound without external tracking using deep learning. Medical image analysis, 2018. \textit{48: p. 187-202.}}\par}\par

{\fontsize{11pt}{13.2pt}\selectfont 63.\tab Prevost, R., et al., \textit{Deep Learning-Based 3D Freehand Ultrasound Reconstruction with Inertial Measurement Units. 2018.}\par}\par

{\fontsize{11pt}{13.2pt}\selectfont 64.\tab Bayat, S., et al., \textit{Investigation of physical phenomena underlying temporal-enhanced ultrasound as a new diagnostic imaging technique: theory and simulations. IEEE transactions on ultrasonics, ferroelectrics, and frequency control, 2017. \textit{65(3): p. 400-410.}}\par}\par

{\fontsize{11pt}{13.2pt}\selectfont 65.\tab Nahlawi, L., et al., \textit{Stochastic modeling of temporal enhanced ultrasound: Impact of temporal properties on prostate cancer characterization. IEEE Transactions on Biomedical Engineering, 2017. \textit{65(8): p. 1798-1809.}}\par}\par

{\fontsize{11pt}{13.2pt}\selectfont 66.\tab Azizi, S., et al., \textit{Detection of prostate cancer using temporal sequences of ultrasound data: a large clinical feasibility study. International journal of computer assisted radiology and surgery, 2016. \textit{11(6): p. 947-956.}}\par}\par

{\fontsize{11pt}{13.2pt}\selectfont 67.\tab Azizi, S., et al., \textit{Transfer learning from RF to B-mode temporal enhanced ultrasound features for prostate cancer detection. International journal of computer assisted radiology and surgery, 2017. \textit{12(7): p. 1111-1121.}}\par}\par

{\fontsize{11pt}{13.2pt}\selectfont 68.\tab Azizi, S., et al., \textit{Toward a real-time system for temporal enhanced ultrasound-guided prostate biopsy. International journal of computer assisted radiology and surgery, 2018. \textit{13(8): p. 1201-1209.}}\par}\par

{\fontsize{11pt}{13.2pt}\selectfont 69.\tab Azizi, S., et al., \textit{Detection and grading of prostate cancer using temporal enhanced ultrasound: combining deep neural networks and tissue mimicking simulations. International journal of computer assisted radiology and surgery, 2017. \textit{12(8): p. 1293-1305.}}\par}\par

{\fontsize{11pt}{13.2pt}\selectfont 70.\tab Azizi, S., et al., \textit{Deep recurrent neural networks for prostate cancer detection: analysis of temporal enhanced ultrasound. IEEE transactions on medical imaging, 2018. \textit{37(12): p. 2695-2703.}}\par}\par

{\fontsize{11pt}{13.2pt}\selectfont 71.\tab Park, J.H., et al. \textit{Automatic cardiac view classification of echocardiogram. in 2007 IEEE 11th International Conference on Computer Vision. 2007. IEEE.}\par}\par

{\fontsize{11pt}{13.2pt}\selectfont 72.\tab Chykeyuk, K., D.A. Clifton, and J.A. Noble. \textit{Feature extraction and wall motion classification of 2D stress echocardiography with relevance vector machines. in 2011 IEEE International Symposium on Biomedical Imaging: From Nano to Macro. 2011. IEEE.}\par}\par

{\fontsize{11pt}{13.2pt}\selectfont 73.\tab Sengupta, P.P., et al., \textit{Cognitive machine-learning algorithm for cardiac imaging: a pilot study for differentiating constrictive pericarditis from restrictive cardiomyopathy. Circulation: Cardiovascular Imaging, 2016. \textit{9(6): p. e004330.}}\par}\par

{\fontsize{11pt}{13.2pt}\selectfont 74.\tab Samad, M.D., et al., \textit{Predicting survival from large echocardiography and electronic health record datasets: optimization with machine learning. JACC: Cardiovascular Imaging, 2018: p. 2641.}\par}\par

{\fontsize{11pt}{13.2pt}\selectfont 75.\tab Madani, A., et al., \textit{Fast and accurate view classification of echocardiograms using deep learning. NPJ digital medicine, 2018. \textit{1(1): p. 1-8.}}\par}\par

{\fontsize{11pt}{13.2pt}\selectfont 76.\tab Madani, A., et al., \textit{Deep echocardiography: data-efficient supervised and semi-supervised deep learning towards automated diagnosis of cardiac disease. NPJ digital medicine, 2018. \textit{1(1): p. 1-11.}}\par}\par

{\fontsize{11pt}{13.2pt}\selectfont 77.\tab Gao, X., et al., \textit{A fused deep learning architecture for viewpoint classification of echocardiography. Information Fusion, 2017. \textit{36: p. 103-113.}}\par}\par

{\fontsize{11pt}{13.2pt}\selectfont 78.\tab Dezaki, F.T., et al., \textit{Deep residual recurrent neural networks for characterisation of cardiac cycle phase from echocardiograms, in Deep Learning in Medical Image Analysis and Multimodal Learning for Clinical Decision Support. 2017, Springer. p. 100-108.}\par}\par

{\fontsize{11pt}{13.2pt}\selectfont 79.\tab Wu, T., et al., \textit{Machine learning for diagnostic ultrasound of triple-negative breast cancer. Breast cancer research and treatment, 2019. \textit{173(2): p. 365-373.}}\par}\par

{\fontsize{11pt}{13.2pt}\selectfont 80.\tab Wu, K., X. Chen, and M. Ding, \textit{Deep learning based classification of focal liver lesions with contrast-enhanced ultrasound. Optik, 2014. \textit{125(15): p. 4057-4063.}}\par}\par

{\fontsize{11pt}{13.2pt}\selectfont 81.\tab Guo, L.-H., et al., \textit{A two-stage multi-view learning framework based computer-aided diagnosis of liver tumors with contrast enhanced ultrasound images. Clinical hemorheology and microcirculation, 2018. \textit{69(3): p. 343-354.}}\par}\par

{\fontsize{11pt}{13.2pt}\selectfont 82.\tab Guo, L., et al. \textit{CEUS-based classification of liver tumors with deep canonical correlation analysis and multi-kernel learning. in 2017 39th Annual International Conference of the IEEE Engineering in Medicine and Biology Society (EMBC). 2017. IEEE.}\par}\par

{\fontsize{11pt}{13.2pt}\selectfont 83.\tab Wildeboer, R.R., et al., \textit{Automated multiparametric localization of prostate cancer based on B-mode, shear-wave elastography, and contrast-enhanced ultrasound radiomics. European radiology, 2020. \textit{30(2): p. 806-815.}}\par}\par

{\fontsize{11pt}{13.2pt}\selectfont 84.\tab Feng, Y., et al., \textit{A deep learning approach for targeted contrast-enhanced ultrasound based prostate cancer detection. IEEE/ACM transactions on computational biology and bioinformatics, 2018. \textit{16(6): p. 1794-1801.}}\par}\par

{\fontsize{11pt}{13.2pt}\selectfont 85.\tab Chen, Y., et al., \textit{Machine-learning-based classification of real-time tissue elastography for hepatic fibrosis in patients with chronic hepatitis B. Computers in biology and medicine, 2017. \textit{89: p. 18-23.}}\par}\par

{\fontsize{11pt}{13.2pt}\selectfont 86.\tab Gatos, I., et al., \textit{A machine-learning algorithm toward color analysis for chronic liver disease classification, employing ultrasound shear wave elastography. Ultrasound in medicine $\&$  biology, 2017. \textit{43(9): p. 1797-1810.}}\par}\par

{\fontsize{11pt}{13.2pt}\selectfont 87.\tab Sehgal, C.M., et al. \textit{Combined Naïve Bayes and logistic regression for quantitative breast sonography. in 2012 IEEE International Ultrasonics Symposium. 2012. IEEE.}\par}\par

{\fontsize{11pt}{13.2pt}\selectfont 88.\tab Gatos, I., et al., \textit{Temporal stability assessment in shear wave elasticity images validated by deep learning neural network for chronic liver disease fibrosis stage assessment. Medical physics, 2019. \textit{46(5): p. 2298-2309.}}\par}\par

{\fontsize{11pt}{13.2pt}\selectfont 89.\tab Zhang, Q., et al., \textit{Deep learning based classification of breast tumors with shear-wave elastography. Ultrasonics, 2016. \textit{72: p. 150-157.}}\par}\par

{\fontsize{11pt}{13.2pt}\selectfont 90.\tab Zhou, Y., et al., \textit{A radiomics approach with CNN for shear-wave elastography breast tumor classification. IEEE Transactions on Biomedical Engineering, 2018. \textit{65(9): p. 1935-1942.}}\par}\par

{\fontsize{11pt}{13.2pt}\selectfont 91.\tab Zhang, Q., et al., \textit{Dual-mode artificially-intelligent diagnosis of breast tumours in shear-wave elastography and B-mode ultrasound using deep polynomial networks. Medical engineering $\&$  physics, 2019. \textit{64: p. 1-6.}}\par}\par

{\fontsize{11pt}{13.2pt}\selectfont 92.\tab Shi, J., et al., \textit{Stacked deep polynomial network based representation learning for tumor classification with small ultrasound image dataset. Neurocomputing, 2016. \textit{194: p. 87-94.}}\par}\par

{\fontsize{11pt}{13.2pt}\selectfont 93.\tab Gao, J., et al., \textit{A Deep Siamese-Based Plantar Fasciitis Classification Method Using Shear Wave Elastography. IEEE Access, 2019. \textit{7: p. 130999-131007.}}\par}\par

{\fontsize{11pt}{13.2pt}\selectfont 94.\tab Zhou, B. and X. Zhang, \textit{Lung mass density analysis using deep neural network and lung ultrasound surface wave elastography. Ultrasonics, 2018. \textit{89: p. 173-177.}}\par}\par

{\fontsize{11pt}{13.2pt}\selectfont 95.\tab Zhou, B., et al., \textit{Predicting lung mass density of patients with interstitial lung disease and healthy subjects using deep neural network and lung ultrasound surface wave elastography. Journal of the Mechanical Behavior of Biomedical Materials, 2020. \textit{104: p. 103682.}}\par}\par

{\fontsize{11pt}{13.2pt}\selectfont 96.\tab Destrempes, F., et al., \textit{Added Value of Quantitative Ultrasound and Machine Learning in BI-RADS 4–5 Assessment of Solid Breast Lesions. Ultrasound in Medicine $\&$  Biology, 2020. \textit{46(2): p. 436-444.}}\par}\par

{\fontsize{11pt}{13.2pt}\selectfont 97.\tab Roy-Cardinal, M.-H., et al., \textit{Assessment of carotid artery plaque components with machine learning classification using homodyned-K parametric maps and elastograms. IEEE transactions on ultrasonics, ferroelectrics, and frequency control, 2018. \textit{66(3): p. 493-504.}}\par}\par

{\fontsize{11pt}{13.2pt}\selectfont 98.\tab Gangeh, M., et al., \textit{Computer aided prognosis for cell death categorization and prediction in vivo using quantitative ultrasound and machine learning techniques. Medical physics, 2016. \textit{43(12): p. 6439-6454.}}\par}\par

{\fontsize{11pt}{13.2pt}\selectfont 99.\tab Ghorayeb, S.R., et al., \textit{Quantitative ultrasound Texture analysis for differentiating preterm from term fetal lungs. Journal of Ultrasound in Medicine, 2017. \textit{36(7): p. 1437-1443.}}\par}\par

{\fontsize{11pt}{13.2pt}\selectfont 100.\tab Bonet‐Carne, E., et al., \textit{Quantitative ultrasound texture analysis of fetal lungs to predict neonatal respiratory morbidity. Ultrasound in Obstetrics $\&$  Gynecology, 2015. \textit{45(4): p. 427-433.}}\par}\par

{\fontsize{11pt}{13.2pt}\selectfont 101.\tab Caxinha, M., et al., \textit{Automatic cataract classification based on ultrasound technique using machine learning: a comparative study. Physics Procedia, 2015. \textit{70: p. 1221-1224.}}\par}\par

{\fontsize{11pt}{13.2pt}\selectfont 102.\tab Tadayyon, H., et al., \textit{A priori prediction of breast tumour response to chemotherapy using quantitative ultrasound imaging and artificial neural networks. Oncotarget, 2019. \textit{10(39): p. 3910.}}\par}\par

{\fontsize{11pt}{13.2pt}\selectfont 103.\tab Feleppa, E.J., et al. \textit{Prostate-cancer imaging using machine-learning classifiers: Potential value for guiding biopsies, targeting therapy, and monitoring treatment. in 2009 IEEE International Ultrasonics Symposium. 2009. IEEE.}\par}\par

{\fontsize{11pt}{13.2pt}\selectfont 104.\tab Byra, M., et al. \textit{Combining Nakagami imaging and convolutional neural network for breast lesion classification. in 2017 IEEE International Ultrasonics Symposium (IUS). 2017. IEEE.}\par}\par

{\fontsize{11pt}{13.2pt}\selectfont 105.\tab Lekadir, K., et al., \textit{A convolutional neural network for automatic characterization of plaque composition in carotid ultrasound. IEEE journal of biomedical and health informatics, 2016. \textit{21(1): p. 48-55.}}\par}\par

{\fontsize{11pt}{13.2pt}\selectfont 106.\tab Sun, Q., et al., \textit{Deep Learning vs. Radiomics for Predicting Axillary Lymph Node Metastasis of Breast Cancer Using Ultrasound Images: Don't Forget the Peritumoral Region. Frontiers in Oncology, 2020. \textit{10: p. 53.}}\par}\par

{\fontsize{11pt}{13.2pt}\selectfont 107.\tab Lei, Y., et al. \textit{High-Resolution Ultrasound Imaging Reconstruction Using Deep Attention Generative Adversarial Networks. in MEDICAL PHYSICS. 2019. WILEY 111 RIVER ST, HOBOKEN 07030-5774, NJ USA.}\par}\par

{\fontsize{11pt}{13.2pt}\selectfont 108.\tab He, X., et al., \textit{Deep attentional GAN-based high-resolution ultrasound imaging. SPIE Medical Imaging. Vol. 11319. 2020: SPIE.}\par}\par

{\fontsize{11pt}{13.2pt}\selectfont 109.\tab Van Sloun, R.J., et al. \textit{Deep learning for super-resolution vascular ultrasound imaging. in ICASSP 2019-2019 IEEE International Conference on Acoustics, Speech and Signal Processing (ICASSP). 2019. IEEE.}\par}\par

{\fontsize{11pt}{13.2pt}\selectfont 110.\tab van Sloun, R.J., et al., \textit{Super-resolution ultrasound localization microscopy through deep learning. arXiv preprint arXiv:1804.07661, 2018.}\par}\par

{\fontsize{11pt}{13.2pt}\selectfont 111.\tab Rangamani, A., et al., \textit{Landmark detection and tracking in ultrasound using a cnn-rnn framework.}\par}\par

{\fontsize{11pt}{13.2pt}\selectfont 112.\tab Gomariz, A., et al. \textit{Siamese networks with location prior for landmark tracking in liver ultrasound sequences. in 2019 IEEE 16th International Symposium on Biomedical Imaging (ISBI 2019). 2019. IEEE.}\par}\par

{\fontsize{11pt}{13.2pt}\selectfont 113.\tab Huang, P., et al., \textit{Attention‐aware fully convolutional neural network with convolutional long short‐term memory network for ultrasound‐based motion tracking. Medical physics, 2019. \textit{46(5): p. 2275-2285.}}\par}\par

{\fontsize{11pt}{13.2pt}\selectfont 114.\tab Huang, P., et al., \textit{2D ultrasound imaging based intra-fraction respiratory motion tracking for abdominal radiation therapy using machine learning. Physics in Medicine $\&$  Biology, 2019. \textit{64(18): p. 185006.}}\par}\par

{\fontsize{11pt}{13.2pt}\selectfont 115.\tab Brawley, O.W., \textit{Prostate cancer epidemiology in the United States. World journal of urology, 2012. \textit{30(2): p. 195-200.}}\par}\par

{\fontsize{11pt}{13.2pt}\selectfont 116.\tab Ghose, S., et al. \textit{A supervised learning framework for automatic prostate segmentation in trans rectal ultrasound images. in International Conference on Advanced Concepts for Intelligent Vision Systems. 2012. Springer.}\par}\par

{\fontsize{11pt}{13.2pt}\selectfont 117.\tab Ghose, S., et al., \textit{Statistical shape and texture model of quadrature phase information for prostate segmentation. International Journal of Computer Assisted Radiology and Surgery, 2012. \textit{7(1): p. 43-55.}}\par}\par

{\fontsize{11pt}{13.2pt}\selectfont 118.\tab Ghose, S., et al., \textit{A survey of prostate segmentation methodologies in ultrasound, magnetic resonance and computed tomography images. Computer methods and programs in biomedicine, 2012. \textit{108(1): p. 262-287.}}\par}\par

{\fontsize{11pt}{13.2pt}\selectfont 119.\tab Ghose, S., et al., \textit{A supervised learning framework of statistical shape and probability priors for automatic prostate segmentation in ultrasound images. Medical image analysis, 2013. \textit{17(6): p. 587-600.}}\par}\par

{\fontsize{11pt}{13.2pt}\selectfont 120.\tab Ghose, S., et al. \textit{Multiple mean models of statistical shape and probability priors for automatic prostate segmentation. in International Workshop on Prostate Cancer Imaging. 2011. Springer.}\par}\par

{\fontsize{11pt}{13.2pt}\selectfont 121.\tab Sahba, F., H.R. Tizhoosh, and M.M. Salama, \textit{Application of reinforcement learning for segmentation of transrectal ultrasound images. BMC medical imaging, 2008. \textit{8(1): p. 8.}}\par}\par

{\fontsize{11pt}{13.2pt}\selectfont 122.\tab Ghavami, N., et al. \textit{Automatic slice segmentation of intraoperative transrectal ultrasound images using convolutional neural networks. in Medical Imaging 2018: Image-Guided Procedures, Robotic Interventions, and Modeling. 2018. International Society for Optics and Photonics.}\par}\par

{\fontsize{11pt}{13.2pt}\selectfont 123.\tab Ghavami, N., et al., \textit{Integration of spatial information in convolutional neural networks for automatic segmentation of intraoperative transrectal ultrasound images. Journal of Medical Imaging, 2018. \textit{6(1): p. 011003.}}\par}\par

{\fontsize{11pt}{13.2pt}\selectfont 124.\tab Lei, Y., et al., \textit{Ultrasound prostate segmentation based on multidirectional deeply supervised V‐Net. Medical physics, 2019. \textit{46(7): p. 3194-3206.}}\par}\par

{\fontsize{11pt}{13.2pt}\selectfont 125.\tab Karimi, D., et al., \textit{Accurate and robust deep learning-based segmentation of the prostate clinical target volume in ultrasound images. Medical image analysis, 2019. \textit{57: p. 186-196.}}\par}\par

{\fontsize{11pt}{13.2pt}\selectfont 126.\tab Wang, Y., et al., \textit{Deep attentive features for prostate segmentation in 3d transrectal ultrasound. IEEE transactions on medical imaging, 2019. \textit{38(12): p. 2768-2778.}}\par}\par

{\fontsize{11pt}{13.2pt}\selectfont 127.\tab Wang, Y., et al. \textit{Deep attentional features for prostate segmentation in ultrasound. in International Conference on Medical Image Computing and Computer-Assisted Intervention. 2018. Springer.}\par}\par

{\fontsize{11pt}{13.2pt}\selectfont 128.\tab Hassanien, A.E., H. Al-Qaheri, and E.-S.A. El-Dahshan, \textit{Prostate boundary detection in ultrasound images using biologically-inspired spiking neural network. Applied Soft Computing, 2011. \textit{11(2): p. 2035-2041.}}\par}\par

{\fontsize{11pt}{13.2pt}\selectfont 129.\tab Yang, X., et al. \textit{Fine-grained recurrent neural networks for automatic prostate segmentation in ultrasound images. in Thirty-First AAAI Conference on Artificial Intelligence. 2017.}\par}\par

{\fontsize{11pt}{13.2pt}\selectfont 130.\tab Yang, X., et al. \textit{Towards automatic semantic segmentation in volumetric ultrasound. in International Conference on Medical Image Computing and Computer-Assisted Intervention. 2017. Springer.}\par}\par

{\fontsize{11pt}{13.2pt}\selectfont 131.\tab Anas, E.M.A., P. Mousavi, and P. Abolmaesumi, \textit{A deep learning approach for real time prostate segmentation in freehand ultrasound guided biopsy. Medical image analysis, 2018. \textit{48: p. 107-116.}}\par}\par

{\fontsize{11pt}{13.2pt}\selectfont 132.\tab Braz, R., et al. \textit{Breast ultrasound images gland segmentation. in 2012 IEEE International Workshop on Machine Learning for Signal Processing. 2012. IEEE.}\par}\par

{\fontsize{11pt}{13.2pt}\selectfont 133.\tab Jiang, P., et al. \textit{Learning-based automatic breast tumor detection and segmentation in ultrasound images. in 2012 9th IEEE International Symposium on Biomedical Imaging (ISBI). 2012. IEEE.}\par}\par

{\fontsize{11pt}{13.2pt}\selectfont 134.\tab Torbati, N., A. Ayatollahi, and A. Kermani. \textit{Ultrasound image segmentation by using a FIR neural network. in 2013 21st Iranian Conference on Electrical Engineering (ICEE). 2013. IEEE.}\par}\par

{\fontsize{11pt}{13.2pt}\selectfont 135.\tab Torbati, N., A. Ayatollahi, and A. Kermani, \textit{An efficient neural network based method for medical image segmentation. Computers in biology and medicine, 2014. \textit{44: p. 76-87.}}\par}\par

{\fontsize{11pt}{13.2pt}\selectfont 136.\tab Huang, Q., et al., \textit{Automatic segmentation of breast lesions for interaction in ultrasonic computer-aided diagnosis. Information Sciences, 2015. \textit{314: p. 293-310.}}\par}\par

{\fontsize{11pt}{13.2pt}\selectfont 137.\tab Daoud, M.I., et al. \textit{Accurate and fully automatic segmentation of breast ultrasound images by combining image boundary and region information. in 2016 IEEE 13th International Symposium on Biomedical Imaging (ISBI). 2016. IEEE.}\par}\par

{\fontsize{11pt}{13.2pt}\selectfont 138.\tab Lei, B., et al., \textit{Segmentation of breast anatomy for automated whole breast ultrasound images with boundary regularized convolutional encoder–decoder network. Neurocomputing, 2018. \textit{321: p. 178-186.}}\par}\par

{\fontsize{11pt}{13.2pt}\selectfont 139.\tab Singh, V.K., et al., \textit{An Efficient Solution for Breast Tumor Segmentation and Classification in Ultrasound Images Using Deep Adversarial Learning. arXiv preprint arXiv:1907.00887, 2019.}\par}\par

{\fontsize{11pt}{13.2pt}\selectfont 140.\tab Hu, Y., et al., \textit{Automatic tumor segmentation in breast ultrasound images using a dilated fully convolutional network combined with an active contour model. Medical physics, 2019. \textit{46(1): p. 215-228.}}\par}\par

{\fontsize{11pt}{13.2pt}\selectfont 141.\tab Yap, M.H., et al., \textit{Breast ultrasound lesions recognition: end-to-end deep learning approaches. Journal of medical imaging, 2018. \textit{6(1): p. 011007.}}\par}\par

{\fontsize{11pt}{13.2pt}\selectfont 142.\tab Shin, S.Y., et al., \textit{Joint weakly and semi-supervised deep learning for localization and classification of masses in breast ultrasound images. IEEE transactions on medical imaging, 2018. \textit{38(3): p. 762-774.}}\par}\par

{\fontsize{11pt}{13.2pt}\selectfont 143.\tab Carneiro, G., J.C. Nascimento, and A. Freitas, \textit{The segmentation of the left ventricle of the heart from ultrasound data using deep learning architectures and derivative-based search methods. IEEE Transactions on Image Processing, 2011. \textit{21(3): p. 968-982.}}\par}\par

{\fontsize{11pt}{13.2pt}\selectfont 144.\tab Smistad, E. and A. Østvik. \textit{2D left ventricle segmentation using deep learning. in 2017 IEEE International Ultrasonics Symposium (IUS). 2017. IEEE.}\par}\par

{\fontsize{11pt}{13.2pt}\selectfont 145.\tab Leclerc, S., et al., \textit{Deep learning for segmentation using an open large-scale dataset in 2D echocardiography. IEEE transactions on medical imaging, 2019. \textit{38(9): p. 2198-2210.}}\par}\par

{\fontsize{11pt}{13.2pt}\selectfont 146.\tab Lempitsky, V., et al. \textit{Random forest classification for automatic delineation of myocardium in real-time 3D echocardiography. in International Conference on Functional Imaging and Modeling of the Heart. 2009. Springer.}\par}\par

{\fontsize{11pt}{13.2pt}\selectfont 147.\tab Dong, S., et al. \textit{A combined multi-scale deep learning and random forests approach for direct left ventricular volumes estimation in 3D echocardiography. in 2016 Computing In Cardiology Conference (Cinc). 2016. IEEE.}\par}\par

{\fontsize{11pt}{13.2pt}\selectfont 148.\tab Dong, S., et al. \textit{A left ventricular segmentation method on 3D echocardiography using deep learning and snake. in 2016 Computing in Cardiology Conference (CinC). 2016. IEEE.}\par}\par

{\fontsize{11pt}{13.2pt}\selectfont 149.\tab Kass, M., A. Witkin, and D. Terzopoulos, \textit{Snakes: Active contour models. International journal of computer vision, 1988. \textit{1(4): p. 321-331.}}\par}\par

{\fontsize{11pt}{13.2pt}\selectfont 150.\tab Oktay, O., et al., \textit{Anatomically constrained neural networks (ACNNs): application to cardiac image enhancement and segmentation. IEEE transactions on medical imaging, 2017. \textit{37(2): p. 384-395.}}\par}\par

{\fontsize{11pt}{13.2pt}\selectfont 151.\tab Dong, S., et al. \textit{VoxelAtlasGAN: 3D left ventricle segmentation on echocardiography with atlas guided generation and voxel-to-voxel discrimination. in International Conference on Medical Image Computing and Computer-Assisted Intervention. 2018. Springer.}\par}\par

{\fontsize{11pt}{13.2pt}\selectfont 152.\tab Salehi, M., et al. \textit{Precise ultrasound bone registration with learning-based segmentation and speed of sound calibration. in International Conference on Medical Image Computing and Computer-Assisted Intervention. 2017. Springer.}\par}\par

{\fontsize{11pt}{13.2pt}\selectfont 153.\tab Hu, Y., et al., \textit{Weakly-supervised convolutional neural networks for multimodal image registration. Medical image analysis, 2018. \textit{49: p. 1-13.}}\par}\par

{\fontsize{11pt}{13.2pt}\selectfont 154.\tab Hu, Y., et al. \textit{Label-driven weakly-supervised learning for multimodal deformable image registration. in 2018 IEEE 15th International Symposium on Biomedical Imaging (ISBI 2018). 2018. IEEE.}\par}\par

{\fontsize{11pt}{13.2pt}\selectfont 155.\tab Sun, Y., et al., \textit{Towards robust ct-ultrasound registration using deep learning methods, in Understanding and Interpreting Machine Learning in Medical Image Computing Applications. 2018, Springer. p. 43-51.}\par}\par

{\fontsize{11pt}{13.2pt}\selectfont 156.\tab Sun, L. and S. Zhang, \textit{Deformable mri-ultrasound registration using 3d convolutional neural network, in Simulation, Image Processing, and Ultrasound Systems for Assisted Diagnosis and Navigation. 2018, Springer. p. 152-158.}\par}\par

{\fontsize{11pt}{13.2pt}\selectfont 157.\tab Zeng, Q., et al., \textit{Weekly supervised convolutional long short-term memory neural networks for MR-TRUS registration. SPIE Medical Imaging. Vol. 11319. 2020: SPIE.}\par}\par

{\fontsize{11pt}{13.2pt}\selectfont 158.\tab Geraldes, A.A. and T.S. Rocha. \textit{A neural network approach for flexible needle tracking in ultrasound images using kalman filter. in 5th IEEE RAS/EMBS International Conference on Biomedical Robotics and Biomechatronics. 2014. IEEE.}\par}\par

{\fontsize{11pt}{13.2pt}\selectfont 159.\tab Beigi, P., et al., \textit{Detection of an invisible needle in ultrasound using a probabilistic SVM and time-domain features. Ultrasonics, 2017. \textit{78: p. 18-22.}}\par}\par

{\fontsize{11pt}{13.2pt}\selectfont 160.\tab Pourtaherian, A., et al., \textit{Robust and semantic needle detection in 3D ultrasound using orthogonal-plane convolutional neural networks. International journal of computer assisted radiology and surgery, 2018. \textit{13(9): p. 1321-1333.}}\par}\par

{\fontsize{11pt}{13.2pt}\selectfont 161.\tab Pourtaherian, A., et al. \textit{Localization of partially visible needles in 3D ultrasound using dilated CNNs. in 2018 IEEE International Ultrasonics Symposium (IUS). 2018. IEEE.}\par}\par

{\fontsize{11pt}{13.2pt}\selectfont 162.\tab Mwikirize, C., J.L. Nosher, and I. Hacihaliloglu, \textit{Convolution neural networks for real-time needle detection and localization in 2D ultrasound. International journal of computer assisted radiology and surgery, 2018. \textit{13(5): p. 647-657.}}\par}\par

{\fontsize{11pt}{13.2pt}\selectfont 163.\tab Mwikirize, C., J.L. Nosher, and I. Hacihaliloglu, \textit{Learning needle tip localization from digital subtraction in 2D ultrasound. International journal of computer assisted radiology and surgery, 2019. \textit{14(6): p. 1017-1026.}}\par}\par

{\fontsize{11pt}{13.2pt}\selectfont 164.\tab Zhang, Y., et al., \textit{Multi-needle Detection in 3D Ultrasound Images Using Unsupervised Order-graph Regularized Sparse Dictionary Learning. IEEE Transactions on Medical Imaging, 2020.}\par}\par

{\fontsize{11pt}{13.2pt}\selectfont 165.\tab Zhang, Y., et al., \textit{Multi-needle Localization with Attention U-Net in US-guided HDR Prostate Brachytherapy. Medical Physics. \textit{n/a(n/a).}}\par}\par

{\fontsize{11pt}{13.2pt}\selectfont 166.\tab Zhang, Y., et al., \textit{Weakly supervised multi-needle detection in 3D ultrasound images with bidirectional convolutional sparse coding. SPIE Medical Imaging. Vol. 11319. 2020: SPIE.}\par}\par

{\fontsize{11pt}{13.2pt}\selectfont 167.\tab Tom, F. and D. Sheet. \textit{Simulating patho-realistic ultrasound images using deep generative networks with adversarial learning. in 2018 IEEE 15th International Symposium on Biomedical Imaging (ISBI 2018). 2018. IEEE.}\par}\par

{\fontsize{11pt}{13.2pt}\selectfont 168.\tab Hu, Y., et al., \textit{Freehand ultrasound image simulation with spatially-conditioned generative adversarial networks, in Molecular imaging, reconstruction and analysis of moving body organs, and stroke imaging and treatment. 2017, Springer. p. 105-115.}\par}\par

{\fontsize{11pt}{13.2pt}\selectfont 169.\tab Fujioka, T., et al., \textit{Breast Ultrasound Image Synthesis using Deep Convolutional Generative Adversarial Networks. Diagnostics, 2019. \textit{9(4): p. 176.}}\par}\par

{\fontsize{11pt}{13.2pt}\selectfont 170.\tab Gangeh, M.J., et al. \textit{Breast tumour visualization using 3D quantitative ultrasound methods. in Medical Imaging 2016: Ultrasonic Imaging and Tomography. 2016. International Society for Optics and Photonics.}\par}\par

{\fontsize{11pt}{13.2pt}\selectfont 171.\tab Mohareri, O., et al. \textit{Multi-parametric 3D quantitative ultrasound vibro-elastography imaging for detecting palpable prostate tumors. in International Conference on Medical Image Computing and Computer-Assisted Intervention. 2014. Springer.}\par}\par

\vspace{\baselineskip}
\setlength{\parskip}{9.96pt}

\printbibliography
\end{document}